\documentclass[twocolumn,10pt]{article}

\usepackage[utf8]{inputenc}
\usepackage[T1]{fontenc}
\usepackage{amsmath,amsfonts,amssymb,amsthm}
\usepackage{graphicx}
\usepackage{xcolor}
\usepackage{tcolorbox}
\usepackage{hyperref}
\usepackage{cite}
\usepackage{authblk}
\usepackage{geometry}
\usepackage{booktabs}
\usepackage{multirow}
\usepackage{tabularx}
\usepackage{url}
\usepackage{algorithm}
\usepackage{algpseudocode}
\usepackage{tikz}
\usetikzlibrary{shapes,arrows,positioning,calc,patterns,decorations.pathreplacing}
\usepackage{colortbl}
\usepackage{pifont}
\usepackage{array}
\usepackage{ragged2e}
\usetikzlibrary{
  positioning,
  arrows.meta,
  fit,
  backgrounds
}

\usepackage{fancyvrb}
\usepackage{listings}
\usepackage{enumitem}
\usepackage{caption}
\usepackage{subcaption}
\usepackage{balance}

\definecolor{tlsblue}{HTML}{1A4D8F}
\definecolor{lightgrayrow}{gray}{0.95}
\definecolor{codegreen}{rgb}{0.0,0.4,0.0}
\definecolor{codegray}{rgb}{0.5,0.5,0.5}

\hypersetup{
    colorlinks=true,
    linkcolor=blue,
    citecolor=blue,
    urlcolor=cyan,
}

\newtheorem{theorem}{Theorem}

\newtheorem{definition}[theorem]{Definition}

\title{\textbf{Merkle Tree Certificate Post-Quantum PKI for Kubernetes and Cloud-Native 5G/B5G Core}}

\author[1]{Lakshya Chopra}
\author[2]{Vipin Kumar Rathi}

\affil[1]{coRAN Labs Private Limited, New Delhi, India}
\affil[2]{Ramanujan College, University of Delhi, New Delhi, India}

\date{}

\begin{document}

\twocolumn[
\begin{@twocolumnfalse}
\maketitle

\begin{abstract}
Post-quantum signature schemes such as ML-DSA-65 produce signatures
of 3,309 bytes and public keys of 1,952 bytes, which are over 50$\times$ larger
than classical Ed25519. In TLS-authenticated environments like
Kubernetes control planes and 5G Core networks, where every
inter-component connection is mutually authenticated, this overhead
compounds across thousands of handshakes per second. Merkle Tree
Certificates (MTC), currently under development at IETF, replace
per-certificate issuer signatures with Merkle inclusion proofs and,
in the landmark mode, eliminate on-wire signatures from certificate
authentication entirely.
We present MTC-based PKI architectures for Kubernetes and 3GPP 5G
Service-Based Architecture. Starting from the infrastructure layer,
we replace the Kubernetes cluster CA with an MTCA deployment that
issues MTC certificates to control plane components, with cosigners
and a DaemonSet-based landmark distributor. Building on this, we
design a certificate lifecycle for 5G Network Functions deployed
against QORE, a post-quantum 5G Core. We implement MTC proof
construction and verification in Go's \texttt{crypto/tls} and
\texttt{crypto/x509} packages.
Our measurements on an Intel i9-12900 show MTC landmark verification
completing in under 2~$\mu$s as compared to 24~$\mu$s for ECDSA
signature verification with no measurable impact on TLS handshake
time. We further propose a 6G-native architecture where the NRF
serves as the MTCA and the SCP as witness cosigner, and discuss
applicability to Non-Terrestrial Networks.

\end{abstract}

\vspace{1em}
\vspace{1.5em}
\end{@twocolumnfalse}
]

\section{Introduction}
\label{sec:intro}

The standardization and deployment of Post-Quantum Cryptographic (PQC) schemes are essential security upgrades being carried out worldwide to mitigate the emerging threat of advanced quantum computers capable of breaking classical cryptographic schemes based on the Discrete Logarithm Problem (DLP) and the Integer Factorization Problem (IFP)~\cite{nist_first_pq}. However, this transition introduces several practical challenges, particularly due to the significantly larger signature and ciphertext sizes of post-quantum schemes compared to classical ECC- and RSA-based constructions.

While efforts to integrate post-quantum cryptography into existing security protocols such as TLS and IPsec have been largely successful, the primary focus has remained on post-quantum key exchange mechanisms, such as ML-KEM and hybrid approaches, which address the \textit{Harvest Now, Decrypt Later} (HNDL) threat. In contrast, the adoption of post-quantum signature schemes has been slower and more complex. This is largely due to the need for modifications in Public Key Infrastructure (PKI), including new trust anchors, updated certificate formats, and the increased bandwidth overhead of post-quantum certificates and signatures.

Experimental evaluations of post-quantum certificates have demonstrated noticeable increases in latency in protocols such as
TLS~\cite{pq-tls-benchmark, pkic-pq-drawbacks}. An ML-DSA-65 signature
alone is 3,309 bytes compared to 64 bytes for Ed25519 a 51$\times$
increase. When accounting for full certificate chains with Certificate
Transparency (CT) Signed Certificate Timestamps (SCTs), the
per-handshake signature overhead grows from approximately 256~bytes
(Ed25519) to over 13,000~bytes (ML-DSA-65). This has motivated the exploration of alternative approaches, such as Merkle Tree Certificates (MTC), which leverage certificate transparency and a network of trusted auditors to reduce certificate bandwidth while maintaining quantum resistance. Browser vendors, including Chrome, have shown interest in deploying MTC within the WebPKI as an alternative to directly replacing classical X.509 certificates with post-quantum variants~\cite{chrome-mtc}.

The work in~\cite{draft-davidben} presents an evolving specification for MTC, outlining key entities such as Certificate Authorities (CAs), cosigners, and log monitors, along with their roles and interactions. It also discusses integration with TLS~1.3 and introduces optimizations such as signatureless verification. However, the draft primarily targets public PKI environments and provides a high-level architectural overview. Prior work~\cite{rathi2025pqibtls} explores identity-based encryption (IBE) for 5G TLS but does not address Merkle tree based transparency or private PKI deployments. To the best of our knowledge, no prior work explores the application of MTC in private PKI deployments, such as cloud-native 5G Core networks.

In this paper, we extend the MTC architecture to private PKI deployments. We design an MTC-based PKI for cloud-native 5G Core,
implement MTC verification in Go's TLS stack, and present both
analytical and empirical evaluation of authentication overhead.

Our \textbf{contributions} are as follows:
\begin{itemize}[leftmargin=*,nosep]
    \item We design an MTC-based PKI architecture for Kubernetes
    and cloud-native 5G Core, mapping MTC roles to Kubernetes
    workloads with the MTCA as a persistent Deployment, mirroring
    cosigners as independent pods, and a DaemonSet-based landmark
    distributor and deploy it against QORE~\cite{qore-arxiv}, a
    post-quantum 5G Core built on free5GC.
    \item We implement MTC proof construction and verification in
    Go's \texttt{crypto/x509} and \texttt{crypto/tls}
    packages, enabling TLS~1.3 handshakes with MTC certificates
    in both standalone and landmark modes.
    \item We define the end-to-end certificate lifecycle for 5G NFs:
    provisioning via Kubernetes Secrets and ServiceAccount tokens,
    issuance through the MTCA's HTTP API, landmark distribution via
    DaemonSet, and revocation by index ranges.
    \item We present both analytical and empirical evaluation of MTC   authentication overhead, measuring sub-2~$\mu$s landmark
    verification and demonstrating approximately 85\% reduction in
    on-wire certificate size compared to post-quantum X.509 chains.

    \item We propose a 6G-native architecture where the NRF serves
    as the MTCA, the SCP as witness cosigner, and the
    \texttt{Nnrf\_NFManagement} API is extended with MTC issuance
    and landmark events.
    \item We compare MTC-TLS against PQ X.509 and IBE-based
    TLS~\cite{rathi2025pqibtls}, covering authentication overhead,
    verification cost, transparency, and revocation.
\end{itemize}

\subsection{Motivation}
\label{sec:motivation}

\subsubsection{Challenges of Post-Quantum Signature Schemes}

The NIST Post-Quantum Cryptography Standardization process~\cite{nist_first_pq} has produced ML-DSA (FIPS~204) ~\cite{nist_mldsa} as the primary digital signature standard. While ML-DSA provides strong security guarantees based on the hardness of the Module Learning With Errors (MLWE) problem, its signature and key sizes are substantially larger than classical alternatives, as shown in Table~\ref{tab:pq-sizes}.

\begin{table}[h]
\centering
\caption{Signature and Key Sizes of PQC vs.\ Classical Schemes}
\label{tab:pq-sizes}
\small
\begin{tabular}{@{}lrrr@{}}
\toprule
\textbf{Scheme} & \textbf{Pub.\ Key} & \textbf{Signature} & \textbf{Sec.\ Level} \\
\midrule
\rowcolor{lightgrayrow}
ECDSA P-256 & 65~B & 64~B & $\sim$128-bit \\
Ed25519 & 32~B & 64~B & $\sim$128-bit \\
\rowcolor{lightgrayrow}
RSA-2048 & 256~B & 256~B & $\sim$112-bit \\
\midrule
ML-DSA-44 & 1,312~B & 2,420~B & NIST~2 \\
\rowcolor{lightgrayrow}
ML-DSA-65 & 1,952~B & 3,309~B & NIST~3 \\
ML-DSA-87 & 2,592~B & 4,627~B & NIST~5 \\
\rowcolor{lightgrayrow}
SLH-DSA-128f & 32~B & 17,088~B & NIST~1 \\
\bottomrule
\end{tabular}
\end{table}

In a TLS~1.3 handshake, the server transmits its certificate chain and
signs the handshake transcript via \texttt{CertificateVerify}. With X.509 certificates carrying ML-DSA-65 signatures, a single leaf certificate's issuer signature balloons from 64~bytes (Ed25519) to 3,309~bytes, and the public key from 32~bytes to 1,952~bytes. For a depth-2 chain with two CT SCTs, the total on-wire certificate overhead reaches approximately 17,500~bytes---a 32$\times$ increase over classical Ed25519 ($\sim$550~bytes).

For hash-based schemes like SLH-DSA (FIPS 205 ~\cite{nist_slhdsa}), the situation is even more amplified, with signatures exceeding 17~KB per certificate. These sizes strain bandwidth budgets, increase handshake latency, and may cause packet fragmentation, particularly over constrained links.

\paragraph{Impact on Telecom.}
The work in Q-RAN~\cite{qran-arxiv} lists several practical challenges that post-quantum deployments face in the telecom domain, including protocol-level packet fragmentation, throttling, suboptimal transport
layer performance, and incompatibilities with middleboxes. These arise from the overhead of post-quantum primitives, which existing TCP, UDP,
and SCTP stacks are not yet sized to accommodate. IP-layer encryption using IPsec has introduced mechanisms to mitigate some of these issues.

\subsubsection{Impact of Post-Quantum Signatures on 5G Core Service Based Architecture}
Post-quantum cryptographic schemes introduce significantly larger signatures and certificates, along with higher computational costs. While these challenges are well understood in general-purpose systems, their impact is particularly pronounced in the 5G Core Service-Based Architecture (SBA). The 3GPP 5G Core uses a Service-Based Architecture (SBA) where all Network Functions communicate over HTTP/2 with mandatory TLS 1.2+ (TS~33.501, TS~29.500)~\cite{3gpp-33501}.

In SBA, network functions communicate using mTLS 1.3 and OAuth 2.0 over short-lived, service-based interactions. Transitioning these mechanisms to their post-quantum counterparts (PQ-mTLS 1.3 and PQ-OAuth 2.0) directly inherits the increased overhead of post-quantum primitives. This is particularly problematic in a high-throughput environment such as the 5G Core, where network functions frequently establish and end connections. Sequential NF-to-NF calls are initiated whenever a subscriber/user connects to the network, for procedures such as UE registration, authentication, policy application, session management, etc. We suspect that this overhead could become a bottleneck if not handled carefully. It might also lead to operators neglecting SBA security or keeping classical certificates, which could lead to several security attacks. 3GPP SA3 is studying PQ migration (TR~33.942). ML-DSA signatures alone would add approximately 10~KB per handshake, pushing SBI latency beyond acceptable bounds for latency-sensitive interfaces.

This motivates the need for mechanisms that preserve post-quantum security while reducing the overhead associated with certificate transmission and signature verification. In this work, we address this challenge by leveraging Merkle Tree Certificates (MTC) to enable more efficient authentication in SBA, thereby facilitating a smoother transition towards post-quantum secure 5G Core deployments.

Table~\ref{tab:5g-scale} quantifies the problem.
\begin{table}[t]
\centering
\caption{5G Core mTLS Scale and Post-Quantum Impact}
\label{tab:5g-scale}
\small
\begin{tabular}{@{}p{2.8cm}rp{2.5cm}@{}}
\toprule
\textbf{Metric} & \textbf{Value} & \textbf{Impact} \\
\midrule
\rowcolor{lightgrayrow}
NFs per cluster & 20--50 & Per-NF certificates \\
SBI calls/sec & 1K--10K & mTLS handshakes \\
\rowcolor{lightgrayrow}
Cert chain (classical) & $\sim$2~KB & Acceptable overhead \\
Cert chain (PQ) & $\sim$17~KB & Latency concern \\
\rowcolor{lightgrayrow}
Control-Plane latency budget & <10 ms & Sensitive to cert size \\
\bottomrule
\end{tabular}
\end{table}

We believe that MTC could potentially address this by replacing per-certificate issuer signatures with Merkle inclusion proofs of logarithmic size. A landmark MTC certificate carries approximately 736~bytes of proof data with \emph{zero signatures}\footnote{However, the certificate does carry public key for subsequent signature verification which does add more overhead, especially in the post-quantum case.}, achieving bandwidth that is even lower than classical Ed25519 setups, regardless of the underlying cryptographic strength.

\subsubsection{Alignment with Zero-Trust Requirements in 5G Networks}

The current specification (TS~33.501) mandates TLS for SBI protection but does not prescribe specific certificate formats~\cite{3gpp-33501}. MTC provides a natural upgrade path: NFs can transition from classical X.509 to MTC standalone certificates immediately, and then to landmark certificates as the infrastructure matures, without modifying the core TLS protocol. The revocation-by-index mechanism further eliminates the need for external CRL/OCSP infrastructure, which is a known bottleneck in carrier-grade deployments.

\section{Background}
\label{sec:background}

\subsection{Public Key Infrastructure}
The internet security infrastructure is built on a baseline level of trust, which is made possible through a system known as Public Key Infrastructure (PKI). PKI ensures that an entity claiming to be a website owner is legitimate, and authenticated, with their domain name bound to a cryptographic signing key. Thus, it ensures that it remains computationally infeasible for any other party to impersonate a website. This is achieved via a system of notaries - one which acts as a public witness to certify and authenticate clients, known as Certificate Authorities (CAs). CAs issue certificate to a website owner after verifying that it has actual control over the domain and is the intended party. On verification, a cryptographic X.509 certificate is granted to the party which binds a digital identity to the domain, comprising a public key, organization, validity, CA name and signature, etc. This is known as an X.509 certificate, and is issued after a website owner makes a Certificate Signing Request (CSR) to a CA. The website owner securely stores the corresponding private key to the public key in the certificate, later using it in TLS connections with clients - for proving possession, authenticity and integrity.

As a consequence of this, the clients must trust a set of Certificate Authorities, which are further subdivided into Global/Root CAs, which means a lot of faith is put into believing the CA will not misuse its power. The root CAs then distribute trust by certifying intermediate CAs, that can then issue end-entity certificates to domains, which reduces signing overhead on themselves. Certificates issued in this manner form an X.509 certificate chain, which goes from the end-entity (leaf certificate) to the intermediate CA certificate and finally to the root CA certificate.
The CAs must exercise caution in delegating trust and granting authenticity, as it is critical for ensuring safety for internet clients and servers.

\subsubsection{TLS 1.3}
TLS ~1.3, as specified in RFC 8446, establishes an encrypted, integrity-protected and authenticated connection via the use of PKI, with modern asymmetric and symmetric ciphers. The authenticity of servers (and optionally clients) is established by sharing their respective X.509 certificate chains, that are issued by a CA trusted by both the peers. The server sends a \texttt{Certificate} message containing one or more DER-encoded certificates, followed by a \texttt{CertificateVerify} message carrying a digital signature over the handshake transcript~\cite{rfc8446}. The relying party verifies the certificate chain up to a trusted root CA and then uses the leaf certificate's public key to verify the \texttt{CertificateVerify} signature.

The key observation is that TLS authentication involves \emph{two distinct signature roles}:
\begin{enumerate}[nosep]
    \item \textbf{Issuer signatures} on certificates (proving the CA issued this certificate).
    \item \textbf{Entity signatures} in \texttt{CertificateVerify} (proving the server possesses the private key).
\end{enumerate}
MTC targets role (1): it replaces issuer signatures with inclusion proofs. Role (2) remains unchanged-the entity's private key still signs the handshake transcript using whatever algorithm the entity key supports (ECDSA, Ed25519, or eventually ML-DSA).

\subsection{Post-Quantum Cryptographic Transition}

NIST has standardized ML-KEM (FIPS~203) ~\cite{nist_mlkem} for key encapsulation and ML-DSA (FIPS~204) ~\cite{nist_mldsa} for digital signatures~\cite{nist_first_pq}. These primitives base their security on lattice hard problems such as Core-SVP, and M-SIS, which are intractable with the best known algorithm. To counter the post-quantum threat, several protocols have begun PQ transition by including these primitives in their key schedule and authentication routines. The most widely used security protocol - TLS 1.3, has several IETF drafts that integrate ML-KEM into the key exchange via hybrid constructions (e.g., \texttt{X25519MLKEM768}), addressing the HNDL threat with manageable overhead (ML-KEM-768 ciphertexts are 1,088~bytes).

For authentication, the path is harder. ML-DSA-65 at NIST Security
Level~3 produces 3,309-byte signatures with 1,952-byte public keys.
A TLS certificate chain of depth~2 with two CT SCTs carries:
\begin{equation}
\text{Signature overhead} = 2 \times 3{,}309 + 2 \times 3{,}309
  = 13{,}236~\text{bytes}
\label{eq:pq-overhead}
\end{equation}
(2 chain signatures + 2 SCT signatures), compared to
$4 \times 64 = 256$~bytes for Ed25519.

The introduction of post-quantum signature schemes to existing WebPKI is a much slower process, as it would require changes to CA public key pairs, upgrade of trust anchors, and support from relying parties.

\subsection{Certificate Transparency}
Certificate Transparency aims to add accountability and transparency to PKI, by introducing a decentralized ledger (public) which tracks certificate issuance. CT thus, tries to mitigate the problem of mis-issued certificates, which may lead to CA scrutinization. It achieves this by the use of append-only logs, which reuses Merkle Tree Construction storing cryptographic hashes of the issued leaf certificates. CT introduces 3 extra parties, which are:
\begin{itemize}
    \item Log operators
    \item Auditors
    \item Monitors
\end{itemize}
The log operator is a designated entity maintaining the append-only, immutable log of issued certificates. Whenever a CA issues a certificate, it requests the log to add that certificate to the tree, for which the log promises that the issued certificate will appear in the log within a defined set of grace period. This promise is cryptographically formalized via the use of Signed Certificate Transparency (SCT), which is a signature made by the log operator that it has been issued, a timestamp of when it will be added, and the server certificate. The final certificate given by the CA to the end-entity comprises the original CA-signed plus the SCT(s) from the log operator(s). The server shares this certificate with the client.
Thus it enhances certificate visibility, since each certificate — whether correctly issued or mis-issued will be visible. Servers routinely scan the CT logs, and if a malicious certificate is caught, the server may issue an alarm, holding the issuing CA accountable.

Auditors are entities that periodically query Certificate Transparency (CT) logs to verify their correct operation. They check that a log is append-only and consistent over time, and that it honors its Signed Certificate Timestamps (SCTs) by ensuring that certificates appear in the log within the log’s declared Maximum Merge Delay (MMD). Auditors and other CT participants may exchange information about log states using gossip protocols (e.g. draft-ietf-trans-gossip-05) to detect log equivocation or misbehavior. 
Monitors are services and crawlers that aim at alerting the websites for misissuance, allowing for speedy and prompt action.

\subsection{Merkle Tree Certificates}

The MTC specification~\cite{draft-davidben} defines a system where a CA operates an \emph{issuance log}-an append-only binary Merkle tree (per RFC~9162~\cite{rfc9162}). Each issued certificate becomes a leaf in this tree. Instead of signing each certificate individually, the CA periodically signs a \emph{checkpoint} over the tree root, and independent \emph{cosigners} (witnesses and mirrors) verify the log's consistency and countersign the checkpoint.

\begin{definition}[MTC Certificate]
An MTC certificate is a valid X.509 certificate where the \texttt{signatureAlgorithm} is set to \texttt{id-alg-mtcProof} and the \texttt{signatureValue} field contains an encoded \texttt{MTCProof} structure rather than a cryptographic signature. The \texttt{serialNumber} field encodes the certificate's position (log index) in the issuance log.
\end{definition}

The \texttt{MTCProof} contains:
\begin{itemize}[nosep]
    \item \texttt{start}, \texttt{end}: the subtree range $[\text{start}, \text{end})$
    \item \texttt{inclusion\_proof}: sibling hashes from leaf to subtree root
    \item \texttt{signatures}: cosigner signatures (empty for landmarks)
\end{itemize}

MTC defines two certificate modes:
\begin{itemize}[nosep]
    \item \textbf{Standalone}: carries cosigner signatures; verifiable by any relying party that trusts the cosigners.
    \item \textbf{Landmark}: carries no signatures; verifiable only by relying parties that have pre-distributed landmark subtree hashes.
\end{itemize}

\begin{definition}[Landmark Certificate]
A landmark certificate is an MTC certificate whose \texttt{MTCProof} contains an empty \texttt{signatures} field. Verification requires the relying party to possess a pre-distributed subtree hash matching the proof's $[\text{start}, \text{end})$ range. Landmark certificates achieve constant-size authentication overhead regardless of the underlying signature scheme.
\end{definition}

\subsection{Related Work}

Rathi et al.~\cite{rathi2025pqibtls} propose post-quantum Identity-Based TLS (IBE-TLS) for 5G SBA and Kubernetes, using lattice-based IBE to eliminate X.509 certificates entirely. Their approach fully eliminates signature and certificate overhead and replaces it with lightweight KEMs. However, IBE introduces a single point of failure in the Private Key Generator (PKG), lacks transparency mechanisms equivalent to CT, and does not address certificate revocation beyond basic identity revocation.

Chrome's MTC deployment plans~\cite{chrome-mtc} target WebPKI with public CAs. Our work differs by focusing on private PKI environments where the CA, cosigners, and relying parties are all within an administrative domain.

Sikeridis et al.~\cite{pq-tls-benchmark} measure the latency impact of
post-quantum signature candidates on TLS~1.3 handshakes, showing that
authentication overhead dominates over key exchange. The PKI
Consortium~\cite{pkic-pq-drawbacks} further discusses the practical
drawbacks of deploying post-quantum certificates in TLS.

\section{5G Core Network}
The 5G Core network (CN) coupled with the 5G Radio Access Network (RAN) enable wireless connectivity for user devices, such as smartphones. The Core network handles critical functions such as user authentication, authorization, location tracking to provide services, and monitoring and actively participating in fulfilling QoS requirements. The core network provides IP-based connectivity, and abilities to manage and record subscriber usage. 
The core network designed in 3GPP's 5G standardization project aims for a modular, and flexible architecture, allowing for enhanced agility and scalability to support 5G's usecases such as IoT, edge communications, uRLLC, etc. Comprising several well-defined network functions, such as AMF, UDM, AUSF, NRF, etc, the 5G CN adopts network function virtualization (NFV) to enable cloud-native deployment and reduce dependence on specialized hardware. NFV also enables flexible programmable network. 
Each NF offers a set of services, that other network functions can access, after sufficient authentication and authorization. These services define the 5G Service-Based Architecture, catering to the cloud-native vision.

The Table ~\ref{tab:5g-nfs} lists representative network functions with their offered services.
\begin{table*}[t]
\centering
\caption{Representative 5G Core network functions and their services}
\label{tab:5g-nfs}
\small
\begin{tabular}{@{}p{3cm}p{8cm}@{}}
\toprule
\textbf{Network Function} & \textbf{Services} \\
\midrule
AMF & Communication, Mobility Management \\
SMF & PDU Session Management \\
AUSF & UE Authentication \\
UDM & Subscriber Data Management \\
PCF & Policy Control \\
NRF & NF Discovery and Registration \\
\bottomrule
\end{tabular}
\end{table*}

\subsection{Cloud Native 5G Core}
Cloud-native refers to a way of designing applications that leverage the advantages of cloud computing—scalability, resilience, and flexibility. It is not concerned with where applications reside (e.g., public/private cloud, edge), but how they are designed, deployed, and maintained. This approach introduces the concept of microservices, which break an application into smaller, functional, and loosely coupled units that communicate with each other to perform tasks and carry out business logic. In practice, this is achieved using containerized workloads or virtual machines, typically orchestrated by systems such as Kubernetes. This architecture improves manageability, scalability, portability, and deployment speed.

As discussed previously, the 5G Core, as defined by 3GPP, follows a cloud-native vision through its use of microservices, separation of user and control planes, and the Service-Based Architecture (SBA), which enables dynamic discovery and independent scaling of Network Functions (NFs).

In this work, we focus on such a cloud-native 5G Core deployment, where NFs are instantiated as containers and orchestrated using Kubernetes.
\subsection{5G Service-Based Architecture}
5G SBA is a design principle where network functions expose their capabilities (services) via a set of well-defined web interfaces that support RESTful APIs over HTTP/2 with JSON encoding. This moves away from telecom-specific protocols like Diameter (used in 4G EPC) which had limitations such as dynamic peer discovery, and was not flexible or interoperable with cloud native software stack such as load balancers, service meshes, etc. The SBA allows for independent scaling of individual network functions without any tight coupling with other functions. The request-response paradigm allows NFs to get the desired services in real-time, where the NF requesting the services offered by the other is referred to as a consumer, and the NF providing them as the producer. Representative service-based interfaces in the 5G core include \texttt{Namf\_Communication, Nsmf\_PDUSession, Nausf\_UEAuthentication, Nudm\_SubscriberDataManagement, Npcf\_PolicyControl}, and \texttt{Nnrf\_NFDiscovery}.
Furthermore, the availability of event exposure, and subscription APIs allow for asynchronous communication, which releases resources when not in need. This is for example used to notify other NFs when policies, UE status, etc, are changed. 
The APIs are commonly defined through a well-defined spec, in the format of OpenAPI, which provides a guarantee that client and servers agree on a common format. 

\subsection{5G SBA Security Mechanisms}
The security of the core network is essential to maintain services, keep user and critical network data safe, and offer uninterrupted services. The core network is secured from external networks (e.g., the internet) via strict separation between user plane and control plane logic, perimeter security and firewalls (with use of protocols and systems such as IPsec and IDS), and controlled entry through the Network Exposure function (NEF). The internal security is, however, equally important, which is provided with dedicated SBA security mechanisms. 
The 3GPP TS 33.501 specifies the security guidelines for SBA, which includes mechanisms for guaranteeing authentication, encryption, authorization, wherein the protocols of mTLS 1.2+ (usually TLS ~1.3) and OAuth 2.0 is used. The mutual authentication paradigm is used to operate as per NIST's established Zero Trust Network principles, which assume that the attacker is already inside the network. The ability of the NRF to grant access or deny to an NF requesting services allow for zero-trust deployments. In this scenario, the NRF acts as the authorization server. Mutual trust is established through use of Public Key Infrastructure, with Inter/Intra-PLMN CAs, which are responsible for certificate lifecycle management, and delegating trust across the subsets of the network via use of sub-CAs/intermediate CAs.

The Table ~\ref{tab:nrf-oauth} lists the NRF's and NFs role in OAuth 2.0, enabling policy-based access to resources. OAuth 2.0 access tokens are short-lived and bound to mTLS certificates, binding token to real NF identities. The identities themselves are short-lived and rotated frequently to avoid stale certificates and reduce blast radius in case of a security breach.

\begin{table*}[h]
\centering
\caption{Mapping of 5G SBA entities to OAuth 2.0 roles}
\label{tab:nrf-oauth}
\small
\begin{tabular}{p{5cm}|p{8cm}}
\toprule
\textbf{OAuth 2.0 Role} & \textbf{5G SBA Entity} \\
\midrule
Resource Owner & Network Function (NF) owning the service/resource \\
Client & Network Function (NF) consuming a service \\
Authorization Server & NRF (issuing access tokens) \\
Resource Server & Network Function (NF) providing the service \\
\bottomrule
\end{tabular}
\end{table*}

\section {Building PKI for the Infrastructure layer: Kubernetes}
Our work revolves around cloud-native 5G Core, where we use Kubernetes as our orchestrator, which itself is subdivided into a control-plane and user-plane. Kubernetes provides the Deployments, StatefulSets, NetworkPolicies and app to app communication interfaces. The Kubernetes subsystem establishes PKI-level foundations from the control-plane, where it mandates a set of trust anchors, and mutual TLS-protected communications. Keeping this in mind, we focus on demonstrating an MTC-based PKI for Kubernetes, which will naturally build trust from the lower layers of the deployment.

\subsection{Kubernetes PKI: Certificates and Mutual TLS}
Kubernetes PKI is divided into multiple logical certificate authorities (CAs), each serving a distinct trust domain. In a typical deployment, the control plane relies on three primary CAs:

\begin{itemize}
    \item etcd CA
    \item API server (or cluster) CA
    \item front-proxy CA
\end{itemize}

Table~\ref{tab:k8s-ca} summarizes the scope and responsibilities of these CAs.
\begin{table*}[t]
\centering
\caption{Scope of Kubernetes Certificate Authorities}
\label{tab:k8s-ca}
\small
\begin{tabular}{lp{5cm}p{6cm}}
\toprule
\textbf{CA} & \textbf{Scope} & \textbf{Used By} \\
\midrule
etcd CA 
& Secures etcd cluster communication (peer and client TLS) 
& etcd servers, API server (as etcd client) \\

API server (cluster) CA 
& Issues client and serving certificates for control plane components 
& API server, kubelet (client), controllers, administrators \\

front-proxy CA 
& Secures API aggregation layer and front-proxy authentication 
& API server (aggregation layer), extension API servers \\
\bottomrule
\end{tabular}
\end{table*}

These authorities, together with the \texttt{kube-controller-manager} acting as a certificate signer, enable mutual TLS authentication across Kubernetes components by provisioning both serving and client certificates. An entity does not directly obtain certificates from a CA; instead, it submits a \texttt{CertificateSigningRequest (CSR)}, which is a first-class Kubernetes resource type. The CSR is directed at a \emph{signer}, which logically partitions a CA by imposing constraints such as permitted subjects, allowed X.509 extensions, key usages, and whether the CA bit may be set. The built-in signers include \texttt{kubernetes.io/kube-apiserver-client}, \texttt{kubernetes.io/kube-apiserver-client-kubelet}, and \texttt{kubernetes.io/kubelet-serving}.

A CSR object comprises a PEM-encoded PKCS\#10 signing request, the requested signer name, and optionally an expiration period. Requests may be approved either automatically or manually; notably, kubelet serving certificate requests require explicit manual approval. Upon approval, the relevant signing controller validates that the signing conditions are met and then issues the certificate. The signed certificate is stored in \texttt{status.certificate}, marking the request as completed, after which the requesting entity can fetch it from the CSR resource. The components store the certificates in their kubeconfig , for example: \texttt{kubelet.conf, controller-manager.conf, scheduler.conf}

Certificates may also be requested by Pods using their ServiceAccountToken (a mechanism which we reuse in our MTCA deployment), this request, additionally, includes the pod name, pod UID, the requested signer, node name, public key, proof of possession (via signature). This is enabled by the \texttt{PodCertificateRequest feature gate}, and with the \texttt{--runtime-config=certificates.k8s.io/
v1beta1/podcertificaterequests=true} flag set on the API server. This certificate request does not have an approval phase, and the certificate can either be issued, denied or get marked as failed. The signing controller, if it has sufficient privileges, may issue the certificate, which may also involve further inspections of the Pod, where for example, it may load the Pod to read the annotations on it.

\subsubsection{User and Administrator Certificates}
Kubernetes has no built-in \texttt{User} object. Instead, user identity is established through client certificates: the Common Name (CN) field maps to the username, and the Organization (O) fields map to group memberships. For example, a certificate with \texttt{CN=alice} and \texttt{O=system:masters} authenticates as user \texttt{alice} with cluster-admin privileges, since \texttt{system:masters} is bound to \texttt{cluster-admin} by default. The initial admin certificate is generated offline by the \texttt{kubeadm init}, which automatically uses the CA key to sign the certificate, this does not use the CSR API. Next, the bootstrapped admin can approve certificate requests for subsequent users, who access the cluster via kubectl (the human-facing side). The kubectl reads the user credentials from their  kubeconfig. Admins may now also rely on using external identity providers, such as KeyCloak, which provide JSON Web tokens (JWTs), that allows user authentication; this is done by a flow known as OpenID Connect (OIDC). OIDC avoids the revocation problem inherent to certificates-tokens expire naturally and can be invalidated at the identity provider-and integrates with existing organizational identity systems (LDAP, SSO). Thus, user certificates or credentials provide authentication, after which RBAC decides the access control and limits of operations. Kubernetes RBAC operates on identities derived from authenticated requests, where usernames and groups are extracted from certificate subjects (CN and O), and authorization policies bind roles to these identities.

Administrator certificates are not structurally different from regular user certificates-the distinction is purely in the group membership encoded in the certificate's subject. This means that anyone with access to the cluster CA key can mint arbitrary admin credentials, making CA key protection critical. 

Normal user certificates are typically issued via the CSR API (using the \texttt{kube-apiserver-client} signer), while bootstrap or infrastructure certificates (e.g., for the API server itself, the controller manager, or the scheduler) are generated directly by \texttt{kubeadm} during cluster initialization.

\begin{figure*}[h]
\centering
\fbox{\parbox{0.92\textwidth}{
\textbf{Certificate Provisioning Flow: kubelet}
\vspace{0.3em}
\begin{enumerate}
    \item Kubelet starts with a bootstrap token (or bootstrap kubeconfig) that grants only the permission to create CSRs.
    \item Kubelet generates a private key locally and constructs a PKCS\#10 request with subject \texttt{CN=system:node:<nodeName>}, \texttt{O=system:nodes}.
    \item Kubelet submits a \texttt{CertificateSigningRequest} resource targeting the \texttt{kubernetes.io/kube-apiserver-client-kubelet} signer.
    \item The \texttt{csrapproving} controller (or an administrator) approves the CSR after verifying node identity.
    \item The \texttt{csrsigning} controller validates the request against the signer's constraints and issues the certificate.
    \item The signed certificate is written to \texttt{status.certificate}; kubelet retrieves it and begins authenticating with the API server over mTLS.
    \item On certificate expiry, kubelet repeats this flow automatically via the \texttt{RotateKubeletClientCertificate} feature gate (enabled by default since v1.19).
\end{enumerate}
}}
\caption{Kubelet client certificate bootstrap and rotation via the CSR API.}
\label{fig:kubelet-cert-flow}
\end{figure*}

\subsubsection{Mutual TLS 1.3}
Every communications in the Kubernetes control-plane is protected using Mutual TLS, which is made possible using the PKI system described above. For instance, API server communicates via etcd, using the client certificate issued to it by the etcd CA. Similarly, kubelet is issued a client certificate and a serving certificate by the cluster CA, using the mechanism mentioned previously. 

Table \ref{tab:k8s-mtls} lists the common mTLS based connections in the control plane. \footnote{Our discussion primarly focuses on the primary control plane mTLS connections, however, additional TLS-secured connections exist (e.g., kube-proxy to API
server, cloud-controller-manager), which follow the same cluster CA
trust chain and would be covered by the MTCA rotation. Pod-to-API-server communication uses ServiceAccount tokens over one-way TLS and is not
affected by MTC certificate issuance.
}

\begin{table*}[ht]
\centering
\caption{Mutual TLS connections in the Kubernetes control plane.}
\label{tab:k8s-mtls}
\begin{tabular}{lllc}
\toprule
\textbf{Initiator (Client)} & \textbf{Responder (Server)} & \textbf{Trust Anchor (CA)} & \textbf{mTLS} \\
\midrule
API server          & etcd                  & etcd CA              & \checkmark \\
API server          & kubelet               & Cluster CA           & \checkmark\textsuperscript{*} \\
kubelet             & API server            & Cluster CA           & \checkmark \\
controller-manager  & API server            & Cluster CA           & \checkmark \\
scheduler           & API server            & Cluster CA           & \checkmark \\
front-proxy         & API server            & Front-proxy CA       & \checkmark \\
kubectl / users     & API server            & Cluster CA           & \checkmark\textsuperscript{\dag} \\
pods                & API server            & Cluster CA           & -- \\
\bottomrule
\end{tabular}

\vspace{0.5em}
\footnotesize
\textsuperscript{*}Kubelet server certificate verification by the API server requires explicit opt-in (\texttt{--kubelet-certificate-authority}); without it, the API server skips verification of the kubelet's serving cert, reducing this to effectively one-way TLS from the kubelet's perspective. \\
\textsuperscript{\dag}Mutual when client certificates are used; one-way TLS when authentication relies on OIDC tokens or bearer tokens. \\
\texttt{--}\;Pods authenticate via service account tokens over one-way TLS, not mTLS.
\end{table*}

\subsection{MTC-based Cluster CA Deployment Architecture}

We describe an MTCA deployment that replaces the Kubernetes cluster CA for issuing certificates to control plane components. The deployment bootstraps MTC from the conventional X.509 infrastructure already present in a \texttt{kubeadm}-initialized cluster. After the standard certificate setup described above, we deploy the MTCA as a Kubernetes \texttt{Deployment} backed by a \texttt{PersistentVolume} for persistent Merkle log storage, exposing its issuance and configuration endpoints over HTTP (Table~\ref{tab:mtca-api}).

Using this deployment, we rotate the API server and kubelet credentials from conventional X.509 to Merkle-tree certificates, effectively replacing the cluster CA. The new trust anchor is the MTCA log ID together with the cosigner public keys, distributed to components via a \texttt{ConfigMap}. To integrate certificate issuance into Kubernetes, we introduce an \texttt{MTC-Controller}, implemented as a cert-manager \texttt{ExternalIssuer} CRD. This controller watches for \texttt{CertificateSigningRequest} resources; on receiving one, it forwards the request to the MTCA's \texttt{/issue-cert} endpoint. The MTCA validates the request against the trust bootstrap chain-it accepts requests from entities presenting a valid client certificate from the original cluster CA. Upon validation, the MTCA appends a leaf to the Merkle tree, the cosigners independently verify and countersign the log checkpoint, and the controller writes the resulting certificate to the CSR resource's \texttt{status.certificate}
field, from which the requesting component retrieves it. MTC certificates inherit the same \texttt{TBSCertificate} structure from X.509, reusing the \texttt{(CN,O} and \texttt{subjectPublicKeyInfo}, with only signature value being reinterpreted. This ensures the same RBAC behaviour as with X.509.

Initially, the MTCA issues certificates only to control plane components-API server, kubelet, controller-manager, and scheduler. Once the control plane operates on MTC certificates, workload pods are onboarded through the same cert-manager integration: pods annotated with \texttt{cert-manager.io/cluster-issuer: "mtca-issuer"} receive MTC certificates via Kubernetes \texttt{Secrets}, with no application-level changes required.

\paragraph{Trust Distribution.} 
We remark that trust distribution is carried out as a slow, and steady process, which over time replaces the original cluster CA, offering a gradual upgrade. This approach considers the cluster's response to MTC, and always offer a fallback mechanism to standard X.509 in case the MTCA deployment fails. A full rollback procedure, re-signing CSRs with the original cluster
CA, remains available as long as the CA key material is preserved. Additionally, since MTCA, in our scenario, operates over a Kubernetes deployment, we introduce static MTCA log ID, tree, and cosigner public key storages as well. This operates in a manner similar to standard Kubernetes PKI. 

\paragraph{Leader Election and High Availability.}
The MTCA is a single point of dependency for certificate issuance, landmark allocation, and log operation. Naturally, this could lead to a very centralized system with a single point of failure; thus, in an order to mitigate this, the MTCA should run as StatefulSet with leader election to ensure PVC failover for the Merkle log. The MTCA's availability requirement is bounded by certificate lifetime: it must recover before the shortest-lived certificate expires. However, even if it is unavailable, the existing landmarks and certificates remain valid for their lifetime, and do not disrupt ongoing TLS connections - only new ones could be if a new certificate is issued but landmark cpuld not be updated due to downtime.

\label{sec:mtca-k8s}

\subsubsection{TLS~1.3 Compatibility}

Since MTC is a certificate format distinct from X.509, Go's \texttt{crypto/tls} does not natively support it in the TLS~1.3 handshake. Thus, we integrated MTC-relevant patches to the TLS library, adding support for MTC certificate transmission and verification-related operations.
An alternative is to embed the MTC inclusion proof in a custom X.509 extension OID-structurally valid, so Go's TLS stack accepts it-but this would render MTC's signature optimization useless as the X.509 signature continues to exist. A sidecar proxy pattern (similar to Istio's envoy model) is another option, where a local proxy terminates MTC-aware mTLS and the component itself communicates over plaintext to localhost; this avoids Go patches entirely but adds operational complexity. A future Go release with native MTC support in the TLS~1.3 certificate message would remove the need for any of these workaround.


\begin{table}[ht]
\centering
\caption{MTCA HTTP API endpoints.}
\label{tab:mtca-api}

{\setlength{\extrarowheight}{15pt}%
\begin{tabular}{p{2cm}p{2cm}p{3cm}}
\toprule
\textbf{Method} & \textbf{Endpoint} & \textbf{Description} \\
\midrule
\texttt{POST} & \texttt{/issue-cert} & Accepts a public key and component identity; appends a leaf to the Merkle tree and returns the MTC certificate. \\
\texttt{GET}  & \texttt{/trust-
config} & Returns the MTCA log ID, cosigner public keys, and the acceptance policy (e.g., require $k$-of-$n$ cosigners). \\
\texttt{GET}  & \texttt{/landmark-
sequence} & Returns the current active landmark subtree hashes. \\
\texttt{POST} & \texttt{/revoke} & Adds a certificate index to the revoked ranges; included in the next landmark update. \\
\bottomrule
\end{tabular}}
\end{table}

\begin{figure*}[ht]
\centering
\fbox{\parbox{0.94\textwidth}{
\textbf{MTC Certificate Issuance Flow in Kubernetes}
\vspace{0.4em}

\textit{Bootstrap phase (one-time):}
\begin{enumerate}
    \item \texttt{kubeadm init} creates the classical cluster CA and issues X.509 certificates to all control plane components.
    \item MTCA is deployed as a \texttt{Deployment} with a \texttt{PersistentVolume} for log storage. It bootstraps trust by accepting the cluster CA as the initial authentication mechanism.
    \item Cosigner pods are deployed in the \texttt{mtc-cosigners} namespace, each with an independent signing key. \texttt{NetworkPolicy} restricts their communication to the MTCA only.
    \item The \texttt{MTC-Controller} (cert-manager \texttt{ExternalIssuer} CRD) is installed and pointed at the MTCA endpoint.
\end{enumerate}

\vspace{0.3em}
\textit{Certificate issuance (per component):}
\begin{enumerate}\setcounter{enumi}{4}
    \item A component (e.g., kubelet) creates a \texttt{CertificateSigningRequest}, authenticating with its existing X.509 client certificate.
    \item The \texttt{MTC-Controller} intercepts the CSR and calls \texttt{POST /issue-cert} on the MTCA with the component's public key and identity.
    \item The MTCA validates the request against the original cluster CA, generates a Merkle tree leaf, and appends it to the log.
    \item Cosigners independently fetch the updated log, verify consistency, and countersign the checkpoint.
    \item The MTCA returns a standalone MTC certificate; the controller writes it into a Kubernetes \texttt{Secret}.
\end{enumerate}

\vspace{0.3em}
\textit{Landmark distribution (continuous):}
\begin{enumerate}\setcounter{enumi}{9}
    \item Every $\sim$10 minutes, the MTCA produces a new landmark batch of subtree hashes covering recently issued certificates.
    \item The landmark distributor \texttt{DaemonSet} fetches and verifies the new hashes, then writes them to \texttt{/var/run/mtc/landmarks.json} on each node.
    \item Pods mount this file read-only. During mTLS~1.3 handshakes, the verifier checks the peer's landmark certificate against the local hash-no signature verification, only SHA-256 lookup.
\end{enumerate}
}}
\caption{End-to-end MTC certificate lifecycle in Kubernetes.}
\label{fig:mtc-k8s-flow}
\end{figure*}

\begin{table*}[ht]
\centering
\caption{Trust relationships in the MTC Kubernetes deployment.}
\label{tab:mtc-trust}
\begin{tabular}{p{3.2cm}p{2.8cm}p{7cm}}
\toprule
\textbf{Entity} & \textbf{Trusts} & \textbf{Mechanism} \\
\midrule
MTCA & Original cluster CA & Validates component CSRs by checking the X.509 client certificate chain. Bootstrap-only; retired once MTC certs are issued. \\
Cosigners & MTCA log integrity & Each cosigner replays the Merkle log independently. Refuses to countersign if entries are missing, reordered, or the log has forked. \\
Components (kubelet, scheduler, etc.) & MTCA + cosigners & Trust anchor: (MTCA log ID, cosigner public keys, acceptance policy). Fetched once via \texttt{GET /trust-config}, stored in a \texttt{ConfigMap}. \\
Verifier (TLS handshake) & Landmark file & Checks peer's landmark cert against \texttt{landmarks.json}. The DaemonSet verifies hashes against cosigner signatures before writing them to disk. \\
etcd & Original etcd CA & \textit{Unchanged.} etcd retains its own separate CA. MTCA log data is on a PVC, not in etcd. Intentional isolation. \\
\bottomrule
\end{tabular}
\end{table*}

\section{MTC Integration in Cloud-Native 5G Core}

We now describe our proposed integration of Merkle Tree Certificates (MTC) into a cloud-native  5G Core. Our design targets future post-quantum upgrades to the 5G Core, enabling efficient authentication mechanisms that remain secure against quantum adversaries. In this context, MTC-based PKI provides significant optimizations by reducing the computational and bandwidth overhead associated with post-quantum certificate schemes.

We adapt the standard, public Internet-oriented MTC infrastructure to a private 5G Core environment, with a particular focus on leveraging the signature-free certificate authentication property of MTC in service-based architecture (SBA) communication.

Our design enables the following properties for SBA:
\begin{itemize}
    \item Signature-free certificate authentication in post-quantum secure mTLS across SBA interfaces, reducing computational overhead during NF-to-NF communication
    \item Reduced impact of large post-quantum certificates on short-lived SBA connections, lowering handshake latency and bandwidth usage
    \item Verifiable certificate transparency through publicly auditable Merkle tree structures
    \item Efficient and Scalable Certificate Revocation Using Compact Revoked Index Ranges
\end{itemize}

We demonstrate this integration using the open-source post-quantum 5G Core \textbf{QORE} \cite{qore-arxiv}, built on free5GC and Aether-SDCore, which extends existing implementations with PQ-TLS 1.3 and PQ-OAuth 2.0 support. An architectural design and feasibility of this approach are presented. This integration does not modify the fundamental behavior of the 5G Core; instead, MTC is introduced as a separate, modular component. 

In subsequent sections, we extend this design to propose native support for MTC in future \textbf{B5G/6G} core architectures, with the goal of informing standardization efforts.

\subsection{QORE Deployment Architectural Description}
QORE is deployed with a set of Kubernetes (K8s) Pods, each running a Network function (e.g., AMF, UPF, NRF). Control plane NFs expose HTTP/2 endpoints using Kubernetes services, mimicking 5G SBA, with each interface being secured using Post-Quantum mutual TLS (PQ-mTLS). The networking is realized using native K8s networking with a CNI plugin (e.g., Calico).

The UPF is deployed as a separate pod, with a high performance datapath, formed using an eBPF-based packet processing pipeline, that intercepts packets at N3 and N6 interfaces, using XDP hooks.

The Kubernetes-based deployment allows for scaling of NFs, greater flexibility, and lifecycle management.

\subsection{MTC-based PKI Architecture Overview}

The MTC-based PKI replaces the existing PQ X.509-based PKI in QORE with MTC-compatible certificates that encode Merkle inclusion proofs and, in standalone mode, cosigner signatures. The design preserves mutual authentication guarantees between Network Functions (NFs), while enabling certificate transparency and signature-free transmission in the landmark mode.

\subsubsection{Components}
A dedicated Kubernetes deployment that runs the MTC PKI subsystem, comprising the MTC CA (MTCA), cosigners, mirrors, and auxiliary services is also introduced. These components provide certificate lifecycle management and introduce trust anchors. This PKI subsystem operates alongside the 5G Core while remaining logically decoupled from NF execution. With the use of MTC, trust is distributed across multiple entities, namely the MTCA and independent cosigners, which reduces single-point of failure risk, increases transparency, and could serve as the first step for decentralization in private 5G cores.

All MTC components are deployed as Kubernetes pods and exposed via cluster services. We reuse the existing mechanisms which issue and manage the NF certificates, while additional services are introduced for MTC-specific functionality. These include:
\begin{itemize}
    \item Landmark distribution services that provide NFs with the latest trusted subtree hashes, enabling signatureless verification
    \item Revocation services that replace CRLs with compact revoked index ranges 
    \item Log access services that expose Merkle tree state and consistency proofs for auditing
\end{itemize}

\subsubsection{Services and APIs}
The MTC PKI subsystem exposes a set of internal services to support certificate lifecycle management and trust distribution. The MTCA provides certificate issuance endpoints, which accept NF certificate requests and return MTC-compatible certificates with inclusion proofs. 

Additional services provide access to trust and transparency data. A landmark distribution service exposes endpoints for retrieving the latest subtree hashes, enabling signature-free verification. A revocation service provides compact revoked index ranges, replacing traditional CRL/OCSP mechanisms. Additionally, log access services expose Merkle   tree state and consistency proofs for auditing and monitoring purposes.

These services are exposed via Kubernetes ClusterIP services and are accessible only within the cluster, ensuring isolation from external networks.The data flow within the system consists of certificate requests from NFs to the MTCA, log updates propagated to cosigners and mirrors, and periodic distribution of landmarks to all nodes. These flows are internal to the cluster and rely on Kubernetes networking for communication.

\subsubsection{Secure Storage and RBAC}
Secure key storage mechanisms from Kubernetes-native PKI deployments are reused, including Secrets Store CSI Driver for mounting key material, and integration with external key management systems via the External Secrets Operator (ESO). Role-Based Access Control (RBAC) policies are defined to ensure that only authorized components (e.g., MTCA and certificate provisioning services) can access or modify key material, while NFs are granted read-only access to their own credentials.

\subsection{Deploying MTC on Kubernetes}
In this section, we describe the components used in MTC, which are MTCA, cosigners, relying parties, and PKI-related services (e.g., certificate revocation, landmark distribution, publishing logs, etc).

\subsubsection{MTC Certificate Authority (MTCA) as the Kubernetes CA}

The Merkle Tree Certificate Authority (MTCA) is responsible for certificate issuance, and maintenance of the append-only log (which is constructed using a Merkle tree). The MTCA, unlike traditional CAs, doesn't sign individual entries, but commits a batch of entries to the log, rendering a checkpoint. This checkpoint is then signed by the cosigner to produce a real standalone certificate which the authenticating party can use in its TLS connections. Note that, we follow the subtree optimization as proposed in the IETF draft, which allows the MTCA to request a signed subtree, resulting in smaller inclusion proofs.

The certificate entry is derived from a \texttt{TBSCertificate}, and stores the content octets of its DER encoding. As specified in ~\cite{draft-davidben}, the entry enum is defined as follows:
\small
\begin{verbatim}
enum {
       null_entry(0), tbs_cert_entry(1), (2^16-1)
} MerkleTreeCertEntryType;
\end{verbatim}
\normalsize

This identifies a certificate validated and certified by the CA. The hashed version of this entry is inserted as a leaf in the log. MTCA routinely produces checkpoints (which are subtrees that start at root), defined by the tuple $\mathrm{(root,size)}$; these represent the root of the Merkle tree and the log state -- the number of entries. 

The MTCA is identified by a unique \emph{log identifier} (log ID), which forms the base trust anchor for the system. This identifier is distributed to relying parties and is used in conjunction with cosigner identifiers to construct trust anchor IDs.

\paragraph{Trust Anchor Identifier.}
MTC defines trust anchor identifiers (TAIDs) ~\cite{draft-tls-trust-anchors} as structured identifiers that encode the trust root of a certificate. These serve to aid the server in choosing the certificate corresponding to a CA/cosigner preferred by the client. These may correspond to:
\begin{itemize}
    \item The log identifier (for standalone certificates)
    \item Cosigner identifiers (for cosigned checkpoints)
    \item Landmark identifiers (for subtree-based verification)
\end{itemize}

In landmark-based operation, trust anchor IDs are derived from a base OID and a landmark index. Periodically, the tree size of the CA's most recent checkpoint is designated as a landmark. This defines a corresponding subtree (i.e., the landmark subtree), which serves as a common reference point for inclusion proofs of certificates issued up to that tree size.

\begin{table}[h]
\centering
\caption{Example Trust Anchor Identifiers}
\begin{tabular}{ll}
\toprule
\textbf{Type} & \textbf{Example} \\
\midrule
Log ID & 32473 \\
Cosigner ID & 32473.1 \\
Landmark TAID & 32473.1.42 \\
\bottomrule
\end{tabular}
\end{table}

Here, \texttt{32473.1.42} represents landmark 42 in a sequence derived from base identifier \texttt{32473.1}, as defined in the IETF draft ~\cite{draft-davidben}.

\paragraph{Certificate Types.}
The MTCA issues two types of certificates:

\begin{itemize}
    \item \textbf{Standalone Certificates}: Contain inclusion proofs and cosignatures, enabling verification without having pre-distributed landmarks and trusted subtrees.
    \item \textbf{Landmark Certificates}: Contain only inclusion proofs relative to a known subtree root (landmark), eliminating the need for signatures.
\end{itemize}

\begin{table}[h]
\centering
\caption{Certificate Types in MTC}
\begin{tabular}{lcc}
\toprule
\textbf{Feature} & \textbf{Standalone} & \textbf{Landmark} \\
\midrule
Inclusion Proof & Yes & Yes \\
Cosignature & Yes & No \\
Landmark Dependency & No & Yes \\
Certificate Size & Larger & Smaller \\
\bottomrule
\end{tabular}
\end{table}

\paragraph{Log Structure.}
The issuance log is a binary Merkle tree where each leaf corresponds to a certificate entry. The leaf hash is computed as:

\[
H_{leaf} = H(\texttt{TBSCertificate})
\]

Internal nodes are computed as:

\[
H_{node} = H(H_{left} \,\|\, H_{right})
\]

The MTCA maintains append-only semantics and provides inclusion and consistency proofs to support verification.

\paragraph{Deployment in QORE.}
We reuse the same deployment for MTCA, as described in Section \ref{sec:mtca-k8s}. The QORE NFs use init containers as sidecars to request certificates upon startup by calling the HTTP API endpoints, exposed by the MTCA. We note, however, that \texttt{cert-manager} configured with the MTCA external issuer can also be used for this.

\begin{table}[h]
\centering
\caption{Simplified TBSCertificate Structure}
\begin{tabular}{ll}
\toprule
\textbf{Field} & \textbf{Description} \\
\midrule
subject & NF identity \\
subjectPublicKeyInfo & Public key \\
validity & Time bounds \\
extensions & Optional metadata \\
\bottomrule
\end{tabular}
\end{table}

\begin{table}[h]
\centering
\caption{Landmark Sequence Parameters}
\begin{tabular}{p{3cm}p{3.5cm}}
\toprule
\textbf{Parameter} & \textbf{Description} \\
\midrule
base\_id & OID base for landmark TAIDs \\
max\_landmarks & Max active landmarks stored by RP \\
landmark\_url & Endpoint for fetching latest landmarks \\
\bottomrule
\end{tabular}
\end{table}

\subsubsection{Cosigners}

The cosigner is a trust anchor introduced by MTC, responsible for signing commitments to the state of the issuance log. These commitments correspond to log checkpoints (i.e., Merkle tree roots at a given size), which represent a consistent snapshot of all issued certificates.

A cosigner verifies that the log evolves in an append-only manner by validating consistency proofs between successive checkpoints. The consistency proofs prove that a checkpoint or a subtree is contained within some other subtree. Specifically, given a previously signed checkpoint and a new checkpoint produced by the MTCA, the cosigner verifies that the new checkpoint is a valid extension of the prior tree. Upon successful verification, the cosigner signs the new checkpoint, producing a \emph{cosignature}.
\textit{For the case of subtree signing, the cosigner must only sign subtrees which are contained within some signed checkpoint. 
}
Cosignatures are embedded in standalone MTC certificates, enabling verification in the absence of locally available landmarks. Each cosigner is identified by a unique cosigner identifier and a corresponding public key, which together form part of the trust anchor configuration distributed to Network Functions (NFs).

The IETF draft allows flexibility in cosigner deployment. Cosigners may operate in a lightweight mode, where only consistency proofs are verified, or in a mirroring mode, where the cosigner maintains a local replica of the log. Mirroring cosigners incur higher storage and bandwidth overhead but provide stronger auditability guarantees.

\paragraph{\textbf{Deployment}}

In our deployment, we utilize mirroring cosigners that maintain a full or partial replica of the log and perform both consistency verification and cosignature generation. The interface between the MTCA and cosigners is restricted to authenticated, internal cluster communication to ensure integrity of log updates. Cosigners run as independent \texttt{Deployments} in a separate namespace (\texttt{mtc-cosigners}), each with its own \texttt{ServiceAccount} and signing key stored in a dedicated \texttt{Secret}. Their isolation is enforced through \texttt{NetworkPolicy}: cosigner pods can communicate only with the MTCA and cannot reach other cluster components or external services\footnote{The IETF draft permits public cosigner interfaces, which clients can use to request subtree signatures from mirrors or witnesses. Authenticating parties can also request cosignatures that the CA did not obtain at certificate issuance.}. Each cosigner independently replays the Merkle log and will refuse to countersign if entries are missing, reordered, or if the log has forked; this behaviour prevents and catches CA misbehaviour, which could occur if the MTCA pod is compromised, ensuring that an attacker cannot produce valid certificates without cosigner cooperation.

The cosigner public keys are distributed to NFs and stored as part of their trust configuration. During TLS (and mTLS) authentication, these keys are used to verify cosignatures in standalone certificates. In landmark-based verification, cosigner signatures are not required for individual certificates, but are instead used to validate checkpoint authenticity.

\begin{tcolorbox}[title=Cosignature Generation Process]
Given:
\begin{itemize}
    \item Previous checkpoint: $(\mathrm{root_{old}, size_{old}})$
    \item New checkpoint from MTCA: $(\mathrm{root_{new}, size_{new})}$
    \item Consistency proof $\pi$
\end{itemize}

Steps:
\begin{itemize}
    \item Verify that $\pi$ proves the tree with root $\mathrm{root_{old}}$ is a prefix of the tree with root $\mathrm{root_{new}}$.
    \item Ensure $\mathrm{size_{new} > size_{old}}$.
    \item If verification succeeds, compute:
   \[
   \sigma = \mathrm{{Sign}_{cosigner}(root_{new} \,\|\, size_{new})}
   \]
    \item Output cosignature $\sigma$.
\end{itemize}

Semantics:
\begin{itemize}
\item The cosignature attests that the log has grown append-only from the previously trusted state.
\item Any fork or rewrite of the log would invalidate the consistency proof.
\end{itemize}
\end{tcolorbox}
A consistency proof demonstrates that a newer Merkle tree is an append-only extension of a previously known tree. It consists of a minimal set of intermediate hashes that allow reconstruction of both the old and new roots, ensuring that no previously issued entries have been modified or removed.

\subsection{Relying Parties}

Relying parties (RPs) are responsible for verifying peer certificates during authentication. In MTC, RPs are configured with a set of trust anchors, which include the log identifier, cosigner identifiers with corresponding public keys, and parameters defining acceptable trust configurations.

RPs may be provisioned with trust anchors via out-of-band mechanisms. During certificate verification, the RP selects a certificate whose trust anchor identifier corresponds to a trusted log or cosigner configuration.

To support landmark-based verification, RPs maintain a set of trusted subtree roots (landmarks). These are derived from a landmark sequence, characterized by parameters such as \texttt{base\_id}, \texttt{max\_landmarks}, and a \texttt{landmark\_url}. The RP stores up to \texttt{max\_landmarks} active subtree roots and uses them to verify inclusion proofs in landmark certificates. The RP checks whether the certificate is included in the advertised subtree by validating the subtree against the subtree root. 

Before \textit{accepting a landmark subtree as trusted}, the RP must, however, ensure that the subtree itself is trustworthy, by establishing that the subtree root is consistent with a valid checkpoint of the log. A valid checkpoint is one that satisfies the RP’s cosigner policy, and whose cosignatures are successfully verified using the configured public keys. If the RP maintains older checkpoints, it additionally verifies a consistency proof chain to ensure append-only evolution of the log. The RP must also verify a valid subtree consistency proof linking the subtree root to the checkpoint. Only such validated subtrees are installed as trusted landmarks. RPs may also enforce freshness requirements on checkpoints, for example by checking timestamps or tree sizes against local policy.

RPs advertise their supported trust anchors using the TLS \texttt{trust\_anchor\_id} extension. This allows the peer to select an appropriate certificate type (standalone or landmark). When landmark support is available, landmark certificates are preferred due to their reduced size and computational overhead. If no suitable landmark is available or verification fails, the RP may fall back to verifying standalone certificates using full inclusion proofs to a trusted checkpoint. The RPs may indicate a range of supported landmarks, which is list of consecutive landmarks the client can accept. This is specified in the draft as:
\begin{verbatim}
  struct {
       TrustAnchorID base;
       uint64 min;
       uint64 max;
   } TrustAnchorRange;
\end{verbatim}

\paragraph{\textbf{RPs in QORE:}} 
In our architecture, the Network functions (e.g., AMF, AUSF, NRF, etc) form the relying parties. These Network functions are slightly modified to add support for:
\begin{itemize}
    \item Merkle Tree Certificate-based TLS authentication -- algorithms for verification of the MTC components, such as inclusion proof, consistency proof, cosignatures, etc.
    \item Advertising TLS trust anchor ID extension.
    \item Addition of new trust anchors -- log ID, cosigner ID, etc.
    \item Dynamic update of landmarks.
\end{itemize}
The trust configuration is provisioned at initialization time using Kubernetes Secrets, while landmarks are dynamically updated via the node-local distribution mechanism.

\subsubsection{Log Mirror}

A mirroring cosigner maintains a full replica of the issuance log and cosigns checkpoints only after verifying that all entries are available and the tree is append-only. The IETF draft distinguishes between consistency-checking cosigners, which are lightweight and verify only consistency proofs, and mirroring cosigners, which store the full log contents. A mirror thus provides a stronger guarantee than a witness: it attests not only to a consistent log history, but also to the durable availability of every entry within the log.

In our deployment, the log mirror runs as a Kubernetes \texttt{StatefulSet} backed by persistent storage and exposes an HTTP-based tile interface derived from the \texttt{tlog-tiles} format. Each tile is an immutable, content-addressed blob that can be cached behind standard HTTP caches or CDN edges. The mirror periodically fetches new entries from the MTCA, reconstructs the Merkle tree locally, and verifies that every node is consistent with the CA's checkpoint before producing a cosignature. This full verification enables monitors to audit the log through the mirror rather than the CA directly, reducing load on the CA and improving transparency. In environments where durable logging is required, relying parties should include at least one mirror cosignature in their acceptance policy (Section~7.3 of~\cite{draft-davidben}).

The mirror also serves as the backend for inclusion proof and consistency proof requests. When an NF or external auditor needs to verify that a particular certificate entry is contained in a given checkpoint, it queries the mirror's tile interface for the necessary intermediate hashes. Because tiles are immutable and cacheable, repeated queries for the same tree region are served from cache, making the system efficient even under high query load.

\begin{table}[h]
\centering
\caption{Mirror Service API Endpoints}
\small
\begin{tabular}{@{}p{3.2cm}p{3.8cm}@{}}
\toprule
\textbf{Endpoint} & \textbf{Description} \\
\midrule
\rowcolor{lightgrayrow}
\texttt{GET /tile/\{L\}/\{N\}} & Fetch tile at level $L$, index $N$ \\
\texttt{GET /checkpoint} & Latest signed checkpoint \\
\rowcolor{lightgrayrow}
\texttt{GET /entry/\{index\}} & Individual log entry \\
\texttt{GET /proof/inclusion} & Subtree inclusion proof \\
\rowcolor{lightgrayrow}
\texttt{GET /proof/consistency} & Consistency proof between checkpoints \\
\bottomrule
\end{tabular}
\end{table}

\subsubsection{Landmark Distributor}

The landmark distributor is an auxiliary service that computes and pushes landmark subtree hashes to relying parties. Landmarks are tree sizes (at the latest cosigned checkpoints) designated periodically by the CA; the subtrees they define are the common reference points between authenticating and relying parties that enable signature-free landmark certificates. The IETF draft specifies that landmarks are allocated using a fixed \texttt{time\_between\_landmarks} interval, with at most \texttt{max\_landmarks} active landmarks at any time (Section~6.3.2 of~\cite{draft-davidben}). Before a relying party accepts a landmark subtree as trusted, it must verify that the subtree is consistent with a reference checkpoint cosigned by a sufficient set of cosigners (Section~7.4 of~\cite{draft-davidben}).

Usually, landmark certificates tied to newly issued landmarks are not immediately usable by relying parties. We address this by introducing this service which distributes recent checkpoints in a trusted manner. Relying parties can use these checkpoints to validate subtrees provided by servers by verifying subtree consistency proofs. Once verified against a trusted checkpoint, such subtrees are installed locally as trusted landmarks and can be reused for subsequent certificate verification.

In our Kubernetes deployment, the landmark distributor runs as a \texttt{DaemonSet}, ensuring every worker node receives landmark updates with minimal latency. Each instance of the distributor performs the following steps:

\begin{enumerate}[nosep]
    \item Watches the MTCA for new checkpoint signatures.
    \item Retrieves the latest landmark sequence from the \texttt{landmark\_url} endpoint.
    \item For each new landmark subtree, fetches a subtree consistency proof from the mirror and verifies it against the reference checkpoint.
    \item Validates that the reference checkpoint carries sufficient cosignatures per the node's trust policy.
    \item Writes the validated subtree hashes to a local file (\texttt{/var/run/mtc/landmarks.json}) readable by NF pods on the same node.
    \item The Landmark distributor additionally distributes the list of revoked indices, which allow NFs to update their local trust configuration.
\end{enumerate}
NF pods mount this file as a read-only volume and reload it on change, ensuring that landmark updates propagate to all relying parties without requiring direct network communication between the distributor and each pod.

Thus, the distributor verifies the recent checkpoints on behalf of the relying parties and distributes to them.

\begin{tcolorbox}
\small
\begin{verbatim}
    
# landmark_url response format (Section 6.3.1)
# <last_landmark> <num_active_landmarks>
# tree_size for landmark last_landmark
# tree_size for landmark last_landmark - 1
# ...
42 25
18500
18200
17900
...
\end{verbatim}
\end{tcolorbox}

\normalsize

\subsubsection{Proof Services}

Two categories of proofs are central to the MTC architecture: inclusion proofs and consistency proofs.

An \emph{inclusion proof} (Section~4.3 of~\cite{draft-davidben}) is a sequence of sibling hashes sufficient to reconstruct a subtree root from a single leaf hash. Inclusion proofs are embedded in every MTC certificate-both standalone and landmark-and are the mechanism by which a relying party verifies that a given certificate entry is contained in a trusted subtree. The proof is ordered bottom-up: the first hash is the deepest sibling (closest to the leaf), and the last hash is the top-level sibling (closest to the subtree root). For a subtree of size $n$, the inclusion proof contains at most $\lceil \log_2 n \rceil$ hashes.

A \emph{consistency proof} (Section~4.4 of~\cite{draft-davidben}) demonstrates that a newer Merkle tree is an append-only extension of a previously known tree. Consistency proofs are used in two contexts: (i)~cosigners verify consistency proofs before cosigning a new checkpoint, ensuring the log has not been forked or rewritten; and (ii)~the landmark distributor verifies subtree consistency proofs to confirm that a landmark subtree is contained within a cosigned checkpoint. The proof contains a minimal set of intermediate hashes that allow reconstruction of both the old and new roots.

Both proof types are served by the log mirror via its tile interface. The MTCA and mirror expose endpoints for requesting proofs parameterized by entry index, subtree range, and checkpoint tree size, enabling external auditors and monitors to independently verify log integrity. These proofs are purely hash-based and require no public-key operations, making verification efficient and suitable for constrained environments.

\subsection{Certificate Lifecycle and Trust Distribution}
\label{sec:cert-lifecycle}

In this section, we describe the end-to-end certificate lifecycle management of MTC certificates within QORE. We reiterate, however, this flow is designed to be applicable to any cloud-native 5G core. 

\subsubsection{Provisioning}

At the NF/pod startup, we ensure the NF is configured with enough trust material, such as MTCA Certificate, enough to contact the MTCA service and authenticate it. The rest of the trust material, for example, log ID, cosigner IDs, are usually fetched dynamically (though could be made static).

\begin{enumerate}[nosep]
    \item The pod's init container requests trust configuration from the MTCA service, receiving the log identifier, cosigner public keys and IDs, and cosigner acceptance policy. The MTCA service is verified using the MTCA X.509 Certificate \footnote{While the MTC log ID and cosigners are used to verify certificate issuance, the MTC still does have a standard trust anchor which is used during certificate provisioning.}, which is configured out-of-band. This is similar to TLS bootstrapping in Kubernetes, where, the issuing party's CA or certificate must also be trusted beforehand. The MTCA service URL can be obtained via Kubernetes service discovery across namespaces (e.g., using the service name and namespace), or by explicitly specifying the service name and port. 
    
    \item  The NF may generate a private key on startup and store it securely in a Kubernetes Secret or an external key vault. Alternatively, it can load a pre-provisioned key via the Secrets Store CSI Driver, which mounts the key into the pod’s filesystem from a software- or hardware-backed vault. The External Secrets Operator may be used to synchronize secrets from external stores into Kubernetes when required.
    \item The landmark distributor's \texttt{DaemonSet} ensures that the local landmarks file (\texttt{/var/run/mtc/landmarks.json}) is available on the node before the NF container starts.
    \item The NF loads its trust configuration, private key, and initial landmark, and requests a certificate from the MTCA, where it shares its common name (e.g., AMF), service names, public key, ServiceAccount token, etc. The connection is made over (PQ) TLS 1.3, where the MTCA service is authenticated.
    \item The MTCA validates the request and NF's claimed identity by verifying the ServiceAccount token with the API server at the control plane, and issues a certificate, based on the provided configuration, and appends it to the log. This process can either be automated or manual. The certificate issuance is described in the upcoming section.
    Thus, the initial PKI setup is based on pre-provisioned trust configuration, and bootstrapping, to ensure a secure startup.

\end{enumerate}

\begin{tcolorbox}[title=NF Pod Provisioning Sequence, colback=blue!3,]
\small
\begin{verbatim}
[Init Container]
  |-- GET /mtca/trust-config
  |     -> {log_id, cosigner_keys[], policy}
  |-- Mount /secrets/nf-private-key (CSI)
  |-- Wait for /var/run/mtc/landmarks.json

[NF Container Start]
  |-- Load trust config + key + landmarks
  |     Body: NFProfile + mtcCertRequest{
  |       subjectPublicKeyInfo: <SPKI DER>,
  |       nfType: amf,
  |       serviceAccountToken: <jwt>,
  |       requestedNames: ["amf.5gc.svc"]
  |     }
  |-- Receive 201 + mtcCertificate (standalone)
  |-- Begin serving TLS with MTC cert
  |-- POST /nnrf-nfm/v1/nf-instances/{nfId}
\end{verbatim}
\end{tcolorbox}
\normalsize

\subsubsection{Issuance}

Certificate issuance follows the process defined in Section~6.2 of the IETF draft. Upon receiving a certificate request during the initial provisioning, the MTCA performs the following:

\begin{enumerate}[nosep]
    \item Validates the NF's identity using the steps described previously.
    \item Constructs a \texttt{TBSCertificateLogEntry} from the request, containing the NF's subject, validity period, subject public key algorithm, and a SHA-256 hash of the full \texttt{SubjectPublicKeyInfo}.
    \item Appends the entry to the issuance log as a new leaf, computing the leaf hash as $H_{\text{leaf}} = \text{SHA-256}(\texttt{0x00} \| \text{MerkleTreeCertEntry})$.
    \item Signs the current checkpoint (tree root and size) with the CA cosigner key, then submits the checkpoint and a consistency proof to each witness cosigner.
    \item Collects cosignatures from a sufficient quorum and constructs a standalone certificate containing the inclusion proof and cosignatures.
    \item Returns the standalone certificate to the NF in the registration response.
\end{enumerate}

The standalone certificate is usable immediately upon receipt. The NF can present it in TLS handshakes to any relying party that trusts the CA and cosigners, without waiting for a landmark to be allocated.

\begin{tcolorbox}[title=NF Bootstrapping and Certificate Provisioning Flow]
\small
\begin{enumerate}
    \item NF pod starts with ServiceAccount identity and MTCA trust anchor provisioned.
    \item The supplementary init container retrieves the trust configuration from the MTCA service/
    \item NF generates or loads the TLS signature private key
    \item NF reads initial landmarks from node-local file, where the hostPath is pre-configured.
    \item NF establishes a (PQ)TLS connection to the MTCA
    \item NF sends a certificate request with its ServiceAccount token
    \item MTCA verifies the token via the Kubernetes API server
    \item MTCA appends the entry to the Merkle log and issues the certificate
    \item NF stores the certificate and initializes TLS for SBA communication
\end{enumerate}
\end{tcolorbox}

\begin{figure*}[t]
\centering
\begin{tikzpicture}[
    >=stealth,
    entity/.style={
        rectangle, draw, thick, rounded corners=2pt,
        minimum width=2.2cm, minimum height=0.8cm,
        align=center, font=\small\bfseries, fill=#1
    },
    arr/.style={->, thick},
    darr/.style={->, thick, dashed},
]

\node[entity=blue!8]    (nf)     at (0,0)     {NF Pod};
\node[entity=orange!10] (mtca)   at (5,0)     {MTCA};
\node[entity=green!8]   (cosign) at (10,0)    {Cosigner};
\node[entity=purple!8]  (mirror) at (14.5,0)  {Mirror};

\foreach \n in {nf,mtca,cosign,mirror}
  \draw[thick, black!30] (\n.south) -- ++(0,-12);

\draw[arr] ([yshift=-1.2cm]nf.south) -- 
  node[above, font=\scriptsize] {\texttt{GET /trust-config}}
  ([yshift=-1.2cm]mtca.south);

\draw[darr] ([yshift=-2.0cm]mtca.south) -- 
  node[above, font=\scriptsize] {\{log\_id, cosigner\_keys, policy\}}
  ([yshift=-2.0cm]nf.south);

\draw[arr] ([yshift=-3.2cm]nf.south) -- 
  node[above, font=\scriptsize] {\texttt{POST /issue-cert} + SA token}
  ([yshift=-3.2cm]mtca.south);

\node[font=\scriptsize\itshape, text=black!50, anchor=west]
  at ([yshift=-4.0cm, xshift=0.2cm]mtca.south)
  {Validate token, append leaf to log};

\draw[arr] ([yshift=-5.0cm]mtca.south) -- 
  node[above, font=\scriptsize] {Checkpoint + consistency proof}
  ([yshift=-5.0cm]cosign.south);

\draw[darr] ([yshift=-5.8cm]cosign.south) -- 
  node[above, font=\scriptsize] {Cosignature $\sigma$}
  ([yshift=-5.8cm]mtca.south);

\draw[arr] ([yshift=-6.6cm]mtca.south) -- 
  node[above, font=\scriptsize] {Log entries (replication)}
  ([yshift=-6.6cm]mirror.south);

\draw[arr, line width=1.5pt] ([yshift=-7.8cm]mtca.south) -- 
  node[above, font=\scriptsize] {Standalone MTC cert (proof + cosigs)}
  ([yshift=-7.8cm]nf.south);

\draw[darr] ([yshift=-9.2cm]nf.south) -- 
  node[above, font=\scriptsize] {\texttt{GET /issue-cert} (landmark)}
  ([yshift=-9.2cm]mtca.south);

\draw[darr, line width=1.5pt] ([yshift=-10.0cm]mtca.south) -- 
  node[above, font=\scriptsize] {Landmark cert (736~B proof, 0 sigs)}
  ([yshift=-10.0cm]nf.south);

\node[font=\footnotesize\bfseries, text=black!40, anchor=east]
  at (-0.8,-1.6) {Provision};
\node[font=\footnotesize\bfseries, text=black!40, anchor=east]
  at (-0.8,-5.4) {Issuance};
\node[font=\footnotesize\bfseries, text=black!40, anchor=east]
  at (-0.8,-7.8) {Response};
\node[font=\footnotesize\bfseries, text=black!40, anchor=east]
  at (-0.8,-9.6) {Landmark};

\end{tikzpicture}
\caption{MTC certificate issuance for a 5G NF. Provisioning and
issuance produce a standalone certificate at startup. The landmark
certificate is pulled after the next landmark allocation.}
\label{fig:mtc-issuance-flow}
\end{figure*}
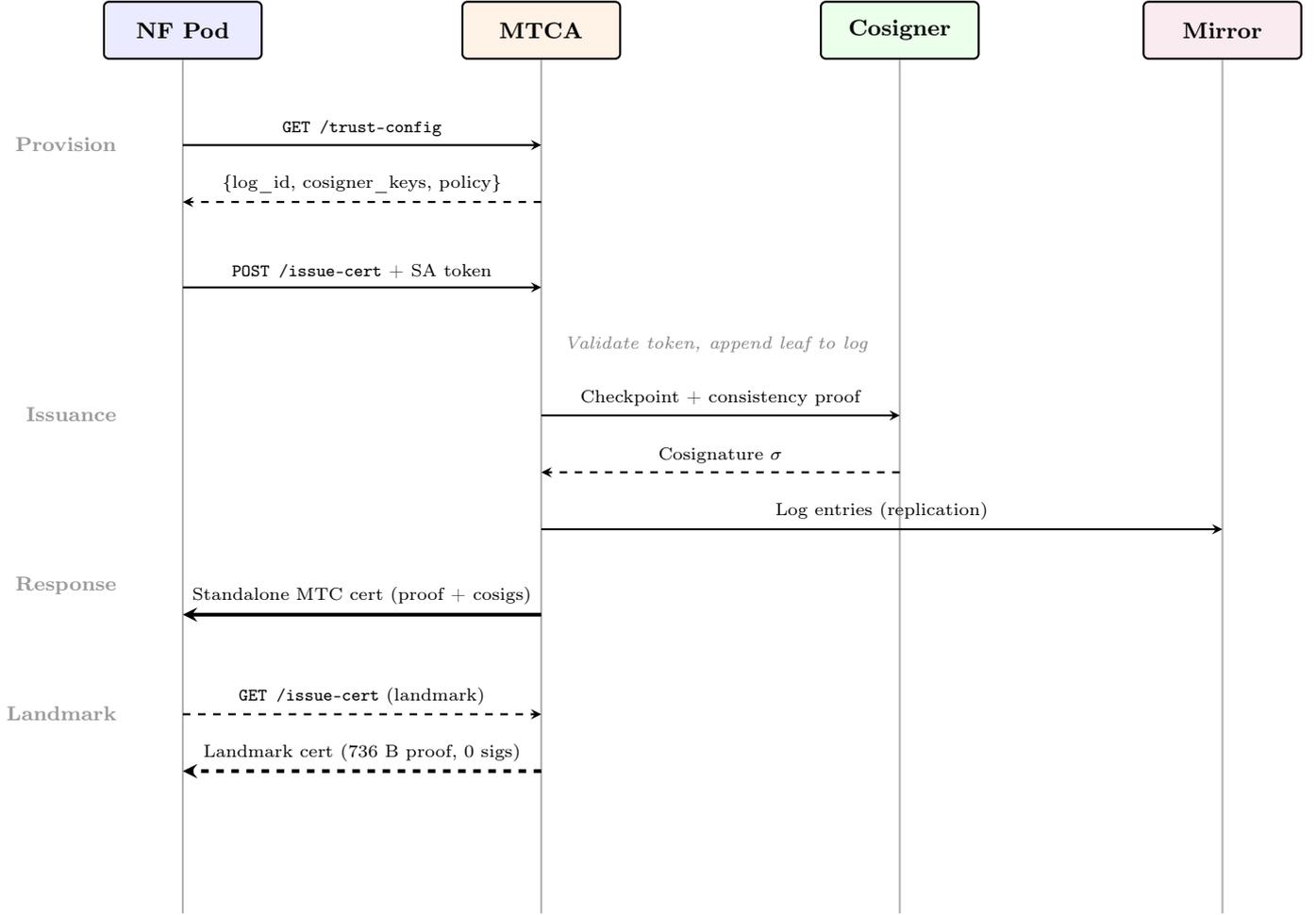

\subsubsection{Landmark Updates}

Landmark certificates become available after the MTCA allocates a new landmark whose subtree contains the NF's log entry. 
The landmark update flow within a Kubernetes-based 5G Core proceeds as follows:

\begin{enumerate}[nosep]
    \item The MTCA allocates a new landmark at tree size $t_{\text{new}}$ and publishes the updated landmark sequence at the \texttt{landmark\_url}.
    \item The MTCA determines the landmark subtrees covering the interval $[t_{\text{prev}}, t_{\text{new}})$ using the subtree decomposition procedure in Section~4.5 of~\cite{draft-davidben}.
    \item For each NF whose log entry falls within a new landmark subtree, the MTCA constructs a landmark certificate with an inclusion proof to that subtree and an empty signatures field.
    \item The landmark distributor \texttt{DaemonSet} on each node fetches the new subtree hashes, verifies subtree consistency proofs against the reference checkpoint, and updates \texttt{/var/run/mtc/landmarks.json}.
        \item NF pods detect the file change, reload their landmark hash set,
    and request their landmark certificate from the MTCA's
    \texttt{/issue-cert} endpoint, specifying the new landmark as
    the target subtree.
    \item The MTCA constructs a landmark certificate with an inclusion
    proof to the relevant subtree and an empty signatures field, and
    returns it to the NF.
    \item The NF stores the landmark certificate alongside its standalone
    certificate and selects between them based on the relying party's
    \texttt{trust\_anchors} extension in the TLS \texttt{ClientHello}.

\end{enumerate}

\begin{figure*}[t]
\centering
\begin{tikzpicture}[
    >=stealth,
    entity/.style={
        rectangle, draw, thick, rounded corners=2pt,
        minimum width=2.2cm, minimum height=0.8cm,
        align=center, font=\small\bfseries, fill=#1
    },
    arr/.style={->, thick},
    darr/.style={->, thick, dashed},
]

\node[entity=orange!10] (mtca)   at (0,0)    {MTCA};
\node[entity=purple!8]  (mirror) at (4.5,0)  {Mirror};
\node[entity=teal!10]   (dist)   at (9,0)    {Landmark\\[-2pt]
    \small Distributor};
\node[entity=blue!8]    (nf)     at (13.5,0) {NF Pod};

\foreach \n in {mtca,mirror,dist,nf}
  \draw[thick, black!30] (\n.south) -- ++(0,-10);

\draw[arr] ([yshift=-1.0cm]mtca.south) --
  node[above, font=\scriptsize] {1.\ New landmark at $t_{\text{new}}$}
  ([yshift=-1.0cm]mirror.south);

\draw[arr] ([yshift=-2.0cm]dist.south) --
  node[above, font=\scriptsize] {2.\ Request subtree hashes}
  ([yshift=-2.0cm]mirror.south);

\draw[darr] ([yshift=-2.8cm]mirror.south) --
  node[above, font=\scriptsize] {3.\ Hashes + consistency proof}
  ([yshift=-2.8cm]dist.south);

\node[font=\scriptsize\itshape, text=black!50, anchor=west]
  at ([yshift=-3.6cm, xshift=0.2cm]dist.south)
  {4.\ Verify against cosigned checkpoint};

\draw[arr, line width=1.5pt] ([yshift=-4.6cm]dist.south) --
  node[above, font=\scriptsize]
  {5.\ Write \texttt{landmarks.json}}
  ([yshift=-4.6cm]nf.south);

\node[font=\scriptsize\itshape, text=black!50, anchor=west]
  at ([yshift=-5.4cm, xshift=-0.4cm]nf.south)
  {6.\ Detect file change, reload hashes};

\draw[arr] ([yshift=-6.4cm]nf.south) --
  node[above, font=\scriptsize]
  {7.\ Request landmark certificate}
  ([yshift=-6.4cm]mtca.south);

\draw[darr, line width=1.5pt] ([yshift=-7.4cm]mtca.south) --
  node[above, font=\scriptsize]
  {8.\ Landmark cert (736~B proof, 0 sigs)}
  ([yshift=-7.4cm]nf.south);

\node[font=\scriptsize\itshape, text=black!50, anchor=west]
  at ([yshift=-8.4cm, xshift=-0.1cm]nf.south)
  {9.\ Serve landmark cert in TLS};

\end{tikzpicture}
\caption{Landmark update propagation. The distributor DaemonSet
verifies subtree hashes via the mirror and writes them locally.
NFs pull their landmark certificate from the MTCA.}
\label{fig:landmark-update-flow}
\end{figure*}
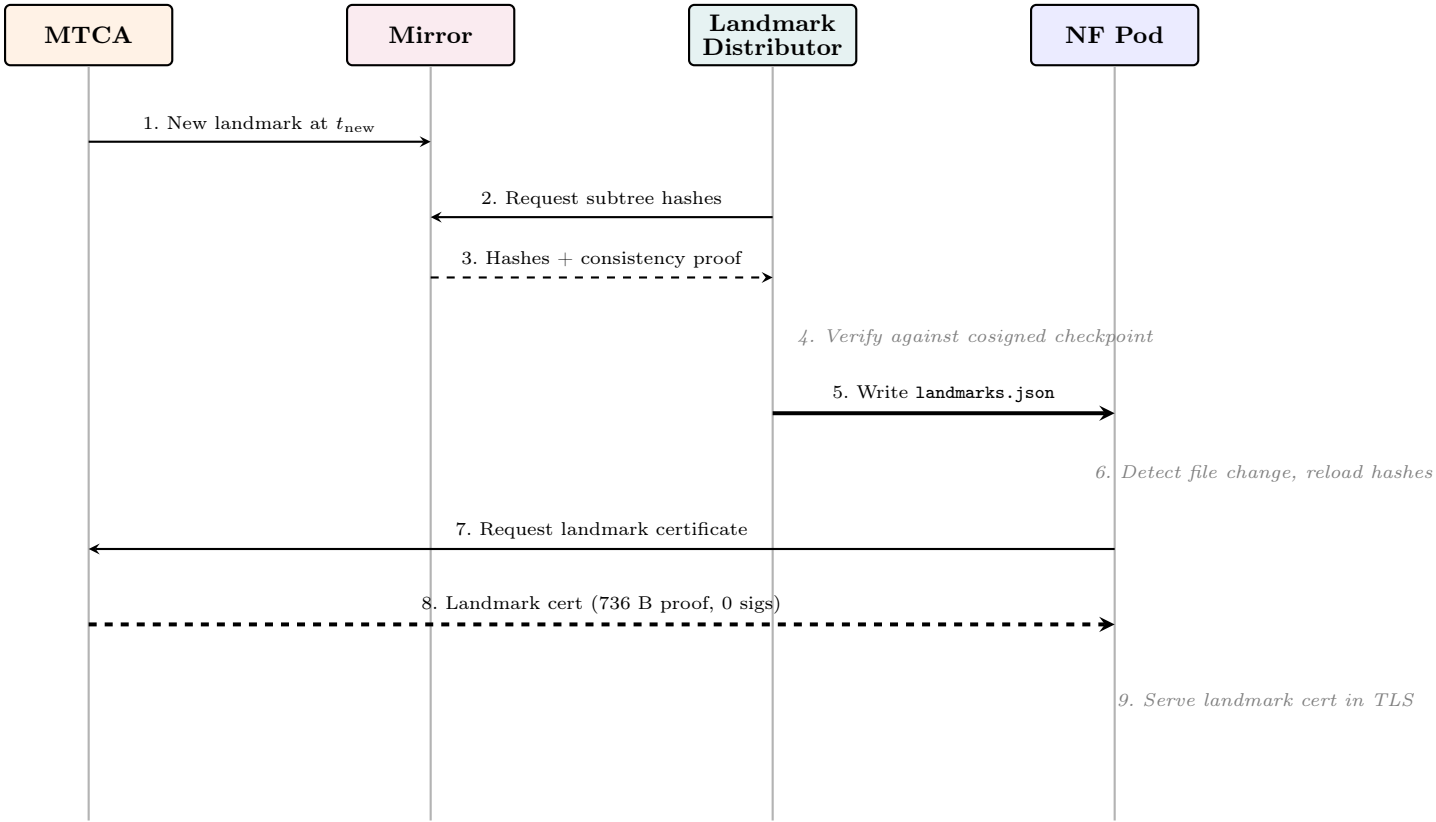

The NF now holds both a standalone and a landmark certificate and selects between them based on the relying party's \texttt{trust\_anchors} extension in the TLS \texttt{ClientHello}.

\subsubsection{Revocation}
\label{sec:revocation-flow}

MTC replaces traditional CRL/OCSP-based revocation with \emph{revocation by index} (Section~7.5 of~\cite{draft-davidben}). Each relying party maintains a list of revoked index ranges $[lo, hi)$ for each trusted log. During certificate verification, the relying party checks whether the certificate's serial number (which equals the log index) falls within any revoked range. This check is a constant-time operation against a sorted list, with no network dependency.

The revocation flow in the 5G Core operates as follows:

\begin{enumerate}[nosep]
    \item The NRF/MTCA determines that a certificate must be revoked (e.g., NF deregistration, key compromise, or policy violation). The system operator can request the MTCA for revocation of a specific certificate.
    \item The MTCA adds the entry's index to the revoked ranges list and pushes the updated list to all NFs alongside the next landmark update via the distributor.
    \item Each NF incorporates the new revoked ranges into its local trust configuration.
    \item Subsequent TLS verification of a certificate whose index falls in $[lo, hi)$ is immediately rejected without evaluating the inclusion proof.
\end{enumerate}

\subsection{MTC-based PQ-TLS in QORE}
\label{sec:mtc-pq-tls}

We now describe the TLS handshake flow when an NF presents an MTC landmark certificate to a peer NF during SBA communication. This flow combines MTC’s signature-free certificate verification with post-quantum key exchange (X25519MLKEM768) and post-quantum signature schemes for CertificateVerify, achieving a handshake where certificate authentication involves no public-key signature verification operations.

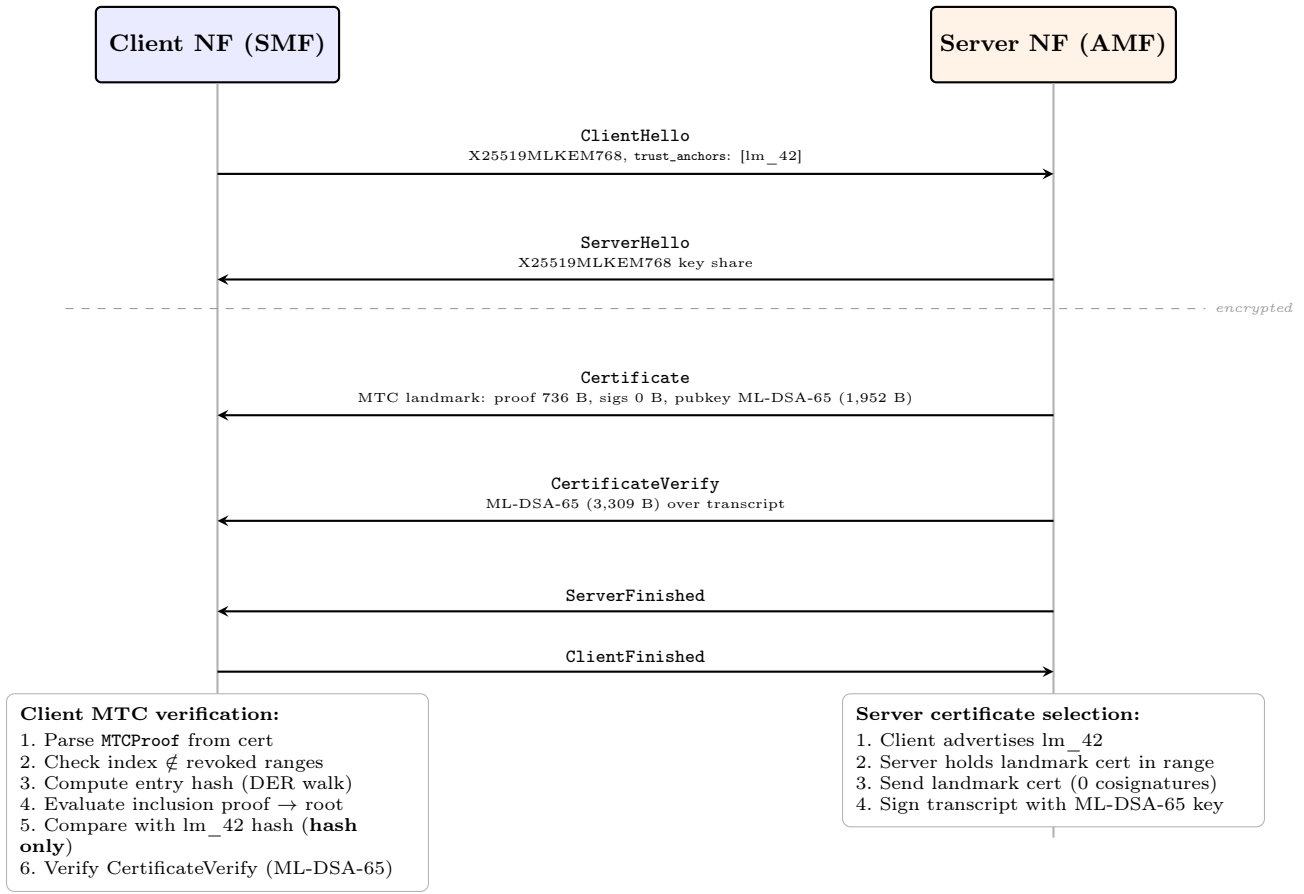
\begin{figure*}[t]
\centering
\begin{tikzpicture}[
    >=stealth,
    entity/.style={
        rectangle, draw, thick, rounded corners=2pt,
        minimum width=3.2cm, minimum height=1.0cm,
        align=center, font=\small\bfseries, fill=#1
    },
    arr/.style={->, thick},
    infobox/.style={
        draw=black!30, rounded corners=3pt,
        inner sep=5pt, font=\scriptsize, align=left,
        text width=5.2cm, fill=white
    },
]

\node[entity=blue!8]    (client) at (0,0)
  {Client NF (SMF)};
\node[entity=orange!10] (server) at (11,0)
  {Server NF (AMF)};

\draw[thick, black!30] (client.south) -- ++(0,-10);
\draw[thick, black!30] (server.south) -- ++(0,-10);

\draw[arr] ([yshift=-1.2cm]client.south) --
  node[above, font=\scriptsize, align=center]
  {\texttt{ClientHello}\\[-1pt]
   {\tiny X25519MLKEM768, \texttt{trust\_anchors}: [lm\_42]}}
  ([yshift=-1.2cm]server.south);

\draw[arr] ([yshift=-2.6cm]server.south) --
  node[above, font=\scriptsize, align=center]
  {\texttt{ServerHello}\\[-1pt]
   {\tiny X25519MLKEM768 key share}}
  ([yshift=-2.6cm]client.south);

\draw[dashed, black!40] (-2,-3.5) -- (13,-3.5)
  node[right, font=\tiny\itshape, text=black!40] {encrypted};

\draw[arr] ([yshift=-4.4cm]server.south) --
  node[above, font=\scriptsize, align=center]
  {\texttt{Certificate}\\[-1pt]
   {\tiny MTC landmark: proof 736~B, sigs 0~B,
    pubkey ML-DSA-65 (1,952~B)}}
  ([yshift=-4.4cm]client.south);

\draw[arr] ([yshift=-5.8cm]server.south) --
  node[above, font=\scriptsize, align=center]
  {\texttt{CertificateVerify}\\[-1pt]
   {\tiny ML-DSA-65 (3,309~B) over transcript}}
  ([yshift=-5.8cm]client.south);

\draw[arr] ([yshift=-7.0cm]server.south) --
  node[above, font=\scriptsize]
  {\texttt{ServerFinished}}
  ([yshift=-7.0cm]client.south);

\draw[arr] ([yshift=-7.8cm]client.south) --
  node[above, font=\scriptsize]
  {\texttt{ClientFinished}}
  ([yshift=-7.8cm]server.south);

\node[infobox, anchor=north] at (0,-8.6) {
  \textbf{Client MTC verification:}\\[2pt]
  1.\ Parse \texttt{MTCProof} from cert\\
  2.\ Check index $\notin$ revoked ranges\\
  3.\ Compute entry hash (DER walk)\\
  4.\ Evaluate inclusion proof $\to$ root\\
  5.\ Compare with lm\_42 hash (\textbf{hash only})\\
  6.\ Verify CertificateVerify (ML-DSA-65)
};

\node[infobox, anchor=north] at (11,-8.6) {
  \textbf{Server certificate selection:}\\[2pt]
  1.\ Client advertises lm\_42\\
  2.\ Server holds landmark cert in range\\
  3.\ Send landmark cert (0 cosignatures)\\
  4.\ Sign transcript with ML-DSA-65 key
};

\end{tikzpicture}
\caption{MTC-based PQ-TLS~1.3 handshake between 5G NFs. Certificate
authentication is hash-only via MTC landmark verification. The sole
PQ signature on the wire is \texttt{CertificateVerify}
(ML-DSA-65, 3,309~B).}
\label{fig:mtc-tls-handshake}
\end{figure*}

The handshake proceeds as follows:

\begin{enumerate}[nosep]
    \item The client NF sends a \texttt{ClientHello} containing an \texttt{X25519MLKEM768} key share for post-quantum key exchange and a \texttt{trust\_anchors} extension listing its latest supported landmark trust anchor ID (e.g., \texttt{landmark\_42}). The client could also instead send a range of supported landmarks. Additionally, support for post-quantum signature schemes, such as ML-DSA-65 is also advertised. 

    \item The server NF responds with \texttt{ServerHello} containing its \texttt{X25519MLKEM768} key share. Both parties derive the handshake secret from the hybrid key exchange, combining X25519 and ML-KEM-768 shared secrets. Specific to the implementation, this may use either X-Wing ~\cite{draft-xwing} or the TLS ECDHE-MLKEM draft ~\cite{draft-ecdhe-mlkem}. The server may decide whether to fallback to standalone certificate or keep using landmark, as per the client's shared data.

    \item The server sends \texttt{EncryptedExtensions}, followed by a \texttt{Certificate} message containing its MTC landmark certificate. This certificate carries an inclusion proof of approximately 736-754~bytes and zero cosigner signatures. The certificate, however, does contain the signature public key for subsequent signature verification.

    \item The server sends \texttt{CertificateVerify}, signing the handshake transcript with its entity private key (e.g., ECDSA~P-256, ML-DSA-65). This proves possession of the private key corresponding to the public key in the certificate.

    \item The client verifies the MTC certificate through the following steps:
    \begin{enumerate}[nosep]
        \item Decodes the \texttt{MTCProof} from the certificate's \texttt{signatureValue} field.
        \item Checks that the serial number (log index) does not appear in any revoked range.
        \item Computes the entry hash by walking the raw DER of the \texttt{TBSCertificate}, skipping the serial number and inner signature algorithm, and replacing the SPKI with its algorithm identifier plus SHA-256 hash.
        \item Evaluates the inclusion proof bottom-up using the procedure in Section~4.3.2 of~\cite{draft-davidben}, obtaining the subtree root hash.
        \item Compares the computed subtree hash against the locally stored hash for \texttt{landmark\_42}. On match, the certificate is accepted with no signature verification.
    \end{enumerate}

    \item The client verifies \texttt{CertificateVerify} using the certificate's public key (standard TLS~1.3 procedure).

    \item Both parties exchange \texttt{Finished} messages, confirming the handshake transcript. The TLS connection is established and HTTP/2 SBI traffic proceeds.
\end{enumerate}

The key observation is that the entire certificate authentication path-from leaf hash to trusted landmark-involves only SHA-256 operations. The only public-key operation in the authentication flow is the \texttt{CertificateVerify} signature, which uses the entity's signature key and is independent of MTC.

\paragraph{Implementation.}
We patch Go's \texttt{crypto/x509} and \texttt{crypto/tls} packages to integrate support for MTC. This patch adds the \texttt{id-alg-mtcProof} OID, an
\texttt{MTCProof} wire format parser, the inclusion proof evaluator from Section~4.3.2 of~\cite{draft-davidben}, DER-based entry hash computation, and revocation-by-index checking. The TLS patch also adds support for \texttt{trust-anchor-id} extensions, allowing clients to specify issuance logs and landmarks. We integrate an MTC proof verification function at the server/client side to verify an MTC certificate, as per the procedure defined in the draft. An extra \texttt{MTCConfig} field provided is in
\texttt{tls.Config} that specifies trusted log IDs, cosigner keys, landmark hashes, and revoked index ranges.

\subsubsection{Comparison with PQ-TLS}

Table~\ref{tab:mtc-vs-pq-tls} provides a qualitative comparison between conventional PQ-TLS (using ML-DSA-65 certificates) and MTC-based PQ-TLS with landmark certificates.

\begin{table}[h]
\centering
\caption{MTC-TLS vs.\ Conventional PQ-TLS (Qualitative)}
\label{tab:mtc-vs-pq-tls}
\small
\begin{tabular}{@{}p{2.3cm}p{2.3cm}p{2.4cm}@{}}
\toprule
\textbf{Property} & \textbf{PQ-TLS} & \textbf{MTC-TLS (Landmark)} \\
\midrule
\rowcolor{lightgrayrow}
Cert auth overhead & $\sim$13.2~KB & $\sim$3~KB \\
Sig.\ verifications & 5 (chain+SCTs) & 0 (hash only) \\
\rowcolor{lightgrayrow}
Key exchange & X25519MLKEM768 & X25519MLKEM768 \\
CertificateVerify & ML-DSA-65 (3.3~KB) & 3.3~KB \textsuperscript{$\ddagger$}\\
\rowcolor{lightgrayrow}
Transparency & Separate CT & Built-in log \\
Revocation & CRL/OCSP & Index ranges \\
\rowcolor{lightgrayrow}
RP state required & Root CA certs & Landmark hashes \\
PQ signature on wire & Yes (multiple) & None (landmark) \\
\bottomrule
\multicolumn{3}{@{}l}{\footnotesize\textsuperscript{$\ddagger$}Same
entity key algorithm; MTC does not change CertificateVerify.}
\end{tabular}
\end{table}

MTC-TLS with landmark certificates eliminates all post-quantum (and classical) signatures from the certificate authentication, where the issuer signatures, SCT signatures, etc., are replaced by hash-based proof. The remaining signature on the wire is the \texttt{CertificateVerify}, which uses the entity's own signature key.
In a post-quantum deployment, this is ML-DSA-65 (3,309~bytes); MTC does
not change the entity key algorithm.

\section{Towards Native MTC in 6G Core Networks}
\label{sec:6g}

In the previous sections, we described an approach for B5G networks to upgrade to an MTC-based PKI stack, where we build from the infrastructure layers, all the way upto the core network. However,  the ongoing development of 6G presents an exciting opportunity to incorporate MTC natively into the core network. With 6G being expected to be fully cloud-native, rely on disaggregated architectures, focus on tighter latency budgets, and expand the usage of Non-Terrestrial Networks, we expect MTC to be a valuable addition. Significantly, the post-quantum cryptographic transition will be a baseline requirement for 6G rather than a retrofit, making the bandwidth and computational costs of post-quantum signatures a first-order design constraint. MTC is a perfect design choice to alleviate post-quantum certificates overhead.

MTC is well-positioned for native 6G integration for several reasons. First, the 6G Service-Based Architecture is expected to retain HTTP/2 (or HTTP/3) with mutual TLS as the baseline inter-function communication model, meaning the certificate overhead problem persists and intensifies with shorter certificate lifetimes and higher NF densities. Second, 6G's emphasis on zero-trust architectures aligns with MTC's built-in transparency guarantees through append-only issuance logs and independent cosigner verification. Third, the landmark optimization-which eliminates all signatures from the certificate authentication path-directly addresses the bandwidth constraints of 6G NTN links, where every byte of handshake overhead impacts latency and throughput.

In this section, we propose a reference architecture for native MTC support in 6G core networks, building on our 5G integration experience and identifying the standardization extensions that would be needed. We begin by creating a role mapping, which links 6G components to their designated MTC roles.

\subsection{Mapping MTC roles to 6G Components}

Table~\ref{tab:5g-roles} maps MTC roles to 6G network functions. The mapping follows the same principles as the 5G integration but accounts for anticipated 6G architectural changes, such as the potential decomposition of the NRF into finer-grained discovery and registration functions.

\begin{table}[t]
\centering
\caption{MTC Role Mapping to 6G Network Functions}
\label{tab:5g-roles}
\small
\begin{tabular}{@{}p{1.8cm}p{2.2cm}p{2.8cm}@{}}
\toprule
\textbf{MTC Role} & \textbf{6G Component} & \textbf{Notes} \\
\midrule
\rowcolor{lightgrayrow}
CA + Log & NRF / Registry Function & Manages NF registration; operates issuance log \\
Witness & SCP & Lightweight; verifies log consistency \\
\rowcolor{lightgrayrow}
Mirror & Dedicated log mirror & Full log copy for transparency and auditing \\
Auth.\ Party & All NFs & AMF, SMF, UPF, AUSF, UDM, etc. \\
\rowcolor{lightgrayrow}
Relying Party & All NFs & Verifies MTC proofs; stores landmarks \\
Monitor & SEPP + operator SOC & Cross-PLMN misissuance detection \\
\bottomrule
\end{tabular}
\end{table}

\subsection{NRF as MTC Certificate Authority}

The NRF is extended by introducing in it, Merkle Tree Certificate Authority (MTCA), that serves to manage the certificates for the entire 6G Core, allowing for lesser dependencies on external PKI systems. Additionally, this provides a built-in security mechanism, with added transparency. The central position of NRF makes it a natural choice for this.

\begin{enumerate}[nosep]
    \item It already knows every NF-NFs register with the NRF (TS~29.510); the NRF can add each NF's certificate to the issuance log at registration time.
    \item It already handles NF profiles (identities, authorizationm discovery), making certificate metadata an incremental addition.
    \item It is (usually) already trusted by all NFs for service discovery.
    \item It already carries high-availability requirements (TS~29.510 Section~5.2), so the issuance log inherits HA properties.
\end{enumerate}

For this purpose, we extend the existing APIs in NRF, introduce new subscription events, and integrate a Merkle-tree based log, to which certificate entries are added.

\subsection{NF Registration with MTC Certificates}
\label{sec:nf-reg}

In the proposed 6G-native architecture, the NRF doubles as the MTC Certificate Authority. Certificate issuance and log management fold into the NF registration workflow-no external PKI subsystem required.

We add one field to the existing \texttt{Nnrf\_NFManagement\_NFRegister} API (TS~29.510): \texttt{mtcCertRequest}, carrying the NF's \texttt{subjectPublicKeyInfo} and its identity attributes (DNS names, PLMN ID, NF type). The NRF validates the NF identity against its profile database, builds a \texttt{TBSCertificateLogEntry}, and appends it to the Merkle issuance log. It then computes an inclusion proof, collects cosignatures for the current checkpoint, and returns a standalone MTC certificate in the registration response. Certificate issuance and NF lifecycle are one operation.

When the NRF allocates a new landmark, it pushes landmark certificates to registered NFs through the existing \texttt{NFStatusNotify} callback, extended with a \texttt{MTC\_LANDMARK\_READY} event type. NFs that receive this notification can switch from standalone to landmark-based verification on their next TLS handshake.

\paragraph{\textbf{NF Registration Response with MTC Certificate (6G)}}
\begin{fbox}
\small
\begin{verbatim}
{
  "nfInstanceId": "smf-456",
  "mtcCertificate": {
    "logId": "32473", "checkpoint": {
      "root": "<hash>",
      "size": 1024
    },
    "inclusionProof": [...],
    "cosignatures": [...]
  }
}
\end{verbatim}
\end{fbox}

\small
\begin{tcolorbox}[title={NF Registration Request with MTC Extension (6G)}]
\begin{verbatim}
    
POST /nnrf-nfm/v1/nf-instances/{nfInstanceId}

{
  "nfType": "SMF",
  "nfStatus": "REGISTERED",
  "ipv4Addresses": ["10.0.0.10"],
  "mtcCertRequest": {
    "subjectPublicKeyInfo": "
        <base64-encoded-key>",
    "dnsNames": ["smf.5gc.local"]
  }
}
\end{verbatim}
\end{tcolorbox}

\normalsize


\begin{tcolorbox}[title=NF Registration with Integrated MTC]
\small
\begin{enumerate}
    \item NF generates key pair
    \item NF sends registration request with mtcCertRequest
    \item NRF validates NF identity
    \item NRF constructs TBSCertificateLogEntry
    \item NRF appends entry to Merkle log
    \item NRF generates checkpoint and inclusion proof
    \item NRF obtains cosignatures (if configured)
    \item NRF returns MTC certificate to NF
    \item NF stores certificate for TLS usage
\end{enumerate}
\end{tcolorbox}

\normalsize

\subsection{SCP as Witness Cosigner}

The Service Communication Proxy (SCP) is a natural candidate for the
witness cosigner role in 6G. The SCP already sits on the data path between
NFs, proxying every SBI request; this position allows it to passively
observe log consistency as a side effect of its normal operation. In 5G,
the SCP is optional (TS~29.500), but 6G architectures are expected to
mandate it for traffic management and observability, making it a
reliably available witness.

The witness design is deliberately lightweight:
\begin{itemize}[nosep]
    \item Each SCP witness stores only the latest checkpoint hash
    (32~bytes of state).
    \item The NRF submits each new checkpoint along with a consistency
    proof from the previous checkpoint.
    \item The SCP verifies the consistency proof and, if valid, returns
    a cosignature.
    \item Deployments require at least 2 SCP witnesses for fault
    tolerance; in multi-site deployments, witnesses should span
    failure domains.
\end{itemize}

The verification overhead is negligible: fewer than 20 SHA-256 operations
per checkpoint. A compromised NF cannot cosign fraudulent checkpoints,
since witness pods run with separate ServiceAccounts and are independently
hardened.

\subsection{Landmark Distribution}
\label{sec:6g-landmark}

We extend the \texttt{Nnrf\_NFManagement\_NFStatusSubscribe} API with a new \texttt{MTC\_LANDMARK\_UPDATE} event type. NFs subscribe during registration, and the NRF pushes landmark updates containing the log ID, landmark number, and subtree hashes. Table~\ref{tab:5g-landmark} summarizes the parameters. This is the core idea taken from 5G's subscription/event exposure APIs.

\begin{table}[t]
\centering
\caption{Landmark Parameters for 6G Core}
\label{tab:5g-landmark}
\small
\begin{tabular}{@{}lrp{3cm}@{}}
\toprule
\textbf{Parameter} & \textbf{Value} & \textbf{Rationale} \\
\midrule
\rowcolor{lightgrayrow}
Cert lifetime & 24~hours & Short-lived for rotation \\
Landmark interval & 10~minutes & Fast availability \\
\rowcolor{lightgrayrow}
max\_landmarks & 145 & $\lceil 24\text{h}/10\text{min} \rceil + 1$ \\
Active subtree hashes & 9,280~B & $290 \times 32$~B per NF \\
\rowcolor{lightgrayrow}
Checkpoint interval & 2~seconds & Low issuance latency \\
\bottomrule
\end{tabular}
\end{table}

\subsection{NF-to-NF mTLS with MTC}

Once NFs hold MTC certificates, the NF-to-NF mTLS flow operates as follows:
\begin{enumerate}[nosep]
    \item The client NF sends a TLS \texttt{ClientHello} including the \texttt{trust\_anchors} extension with its latest landmark trust anchor ID.
    \item The server NF checks whether it holds a landmark certificate whose subtree range matches the client's trust anchor.
    \begin{itemize}[nosep]
        \item \textbf{Match:} Server sends the landmark certificate ($\sim$736~B proof, zero signatures).
        \item \textbf{No match:} Server sends the standalone certificate (proof + cosignatures).
    \end{itemize}
    \item The server sends \texttt{CertificateVerify} (standard classical/PQ signature over handshake transcript).
    \item The client verifies the MTC proof against its landmark hash or cosigner policy, then verifies \texttt{CertificateVerify} using the certificate's public key.
    \item mTLS is established; HTTP/2 SBI traffic proceeds.
\end{enumerate}

\subsection{Certificate Revocation}

MTC replaces CRL/OCSP with \textbf{revocation-by-index} (Section~7.5 of~\cite{draft-davidben}): the NRF maintains a list of revoked index ranges (e.g., $[4200, 4210)$) and pushes updates alongside landmark updates. Each NF checks the local revocation list during proof verification-a constant-time operation with no network dependency. This could be valuable as it avoids OCSP round-trips, and CRL queries. With the use short certificate lifetimes (e.g., 24h), coupled with revocation-by-index, a strong security mechanism with minimal overhead can be devised.

\subsection{Roaming and SEPP Integration}

For inter-PLMN communication, each SEPP is configured with MTC trust
material for its own PLMN's NRF as well as each roaming partner's NRF,
typically provisioned via out-of-band mechanisms during roaming agreement
setup. Landmark hashes are exchanged during SEPP peering (N32-c
signaling), enabling signatureless certificate verification across PLMN
boundaries. SEPPs additionally serve as log monitors, watching partner
logs for unauthorized NF certificate issuance. However, challenges do exist, which we discuss in upcoming sections.

\section{Evaluation and Results}
\label{sec:results}

We evaluate the authentication overhead of MTC-based TLS, and compare it to conventional post-quantum X.509 certificates. We use the methodology which combines theoretical size derivations from the X.509 and MTC wire formats specifications with empirical results. The primary metric is \emph{per-handshake authentication overhead}: the total bytes transmitted for certificate exchange and verification, including certificate chains, signatures, SCTs, and Merkle inclusion proofs where applicable.

\subsection{Certificate Size Analysis}

Table~\ref{tab:cert-sizes} compares per-handshake authentication overhead
across certificate constructions. We separate two components: the
\emph{authentication overhead} (signatures, inclusion proofs, SCTs-the
variable cost that differs between schemes) and the \emph{total on-wire
certificate size} (including the entity public key, subject, and
validity fields). The distinction matters because MTC eliminates
signatures but not the entity public key, which for ML-DSA-65 is
1,952~bytes.

\begin{table*}[t]
\centering
\caption{Per-Handshake Certificate Size Comparison}
\label{tab:cert-sizes}
\small
\begin{tabular}{@{}p{4.2cm}rrrr@{}}
\toprule
\textbf{Scenario} & \textbf{Auth Overhead}
    & \textbf{Entity Pubkey} & \textbf{Base Cert}
    & \textbf{Total On-Wire} \\
\midrule
\rowcolor{lightgrayrow}
X.509 + 2 SCTs (ECDSA P-256)
    & 256~B & 65~B & $\sim$200~B & $\sim$585~B \\
X.509 + 2 SCTs (Ed25519)
    & 256~B & 32~B & $\sim$200~B & $\sim$552~B \\
\rowcolor{lightgrayrow}
X.509 + 2 SCTs (ML-DSA-65)
    & 13,236~B & 1,952~B & $\sim$200~B
    & $\sim$17,504~B \\
\midrule
MTC Standalone (Ed25519 $\times$2)
    & 530~B & 32~B & $\sim$200~B & $\sim$762~B \\
\rowcolor{lightgrayrow}
MTC Standalone (ML-DSA-65 $\times$2)
    & 7,020~B & 1,952~B & $\sim$200~B
    & $\sim$9,172~B \\
MTC Landmark (ECDSA P-256 entity)
    & 754~B & 65~B & $\sim$200~B
    & $\sim$\textbf{1,019~B} \\
\rowcolor{lightgrayrow}
MTC Landmark (ML-DSA-65 entity)
    & 754~B & 1,952~B & $\sim$200~B
    & $\sim$\textbf{2,906~B} \\
\bottomrule
\end{tabular}

\vspace{0.4em}
\footnotesize
Auth overhead = inclusion proof + cosignatures + SCTs (scheme-dependent).
Base cert = subject, validity, extensions, DER framing ($\sim$200~B).
Entity pubkey carried in all cases for \texttt{CertificateVerify}.
\end{table*}

\subsection{Bandwidth Reduction}

Table~\ref{tab:bandwidth-reduction} summarizes the bandwidth reduction achieved by MTC landmark certificates compared to conventional approaches.

\begin{table*}[t]
\centering
\caption{Bandwidth Reduction: MTC Landmark vs.\ Baselines (Total On-Wire)}
\label{tab:bandwidth-reduction}
\small
\begin{tabular}{@{}p{4.5cm}rrr@{}}
\toprule
\textbf{Comparison} & \textbf{Baseline} & \textbf{MTC Landmark}
    & \textbf{Reduction} \\
\midrule
\rowcolor{lightgrayrow}
vs.\ PQ X.509 (ML-DSA-65)
    & $\sim$17,500~B & $\sim$2,900~B & $\sim$\textbf{83\%} \\
vs.\ PQ X.509 (SLH-DSA-128f)
    & $>$36,000~B & $\sim$2,900~B & $>$\textbf{92\%} \\
\rowcolor{lightgrayrow}
vs.\ Classical X.509 (ECDSA P-256)
    & $\sim$585~B & $\sim$1,020~B
    & $-$74\%\textsuperscript{$\dagger$} \\
vs.\ Classical X.509 (Ed25519)
    & $\sim$550~B & $\sim$1,020~B
    & $-$85\%\textsuperscript{$\dagger$} \\
\rowcolor{lightgrayrow}
vs.\ IBE-TLS~\cite{rathi2025pqibtls}
    & $\sim$5,000~B & $\sim$2,900~B & $\sim$\textbf{42\%} \\
\bottomrule
\multicolumn{4}{@{}l}{\footnotesize\textsuperscript{$\dagger$}MTC landmark
with classical entity key is larger than classical X.509; PQ settings
are the target use case.}
\end{tabular}
\end{table*}

MTC landmark certificates provide near constant-size authentication overhead, decoupling certificate transmission cost from the signature scheme. In post-quantum settings, this yields substantial reductions. In classical settings, MTC is slightly larger than Ed25519 but provides built-in transparency and revocation capabilities that traditional X.509 lacks.

\subsection{Empirical Measurements}
\label{sec:empirical}

We measure MTC verification and TLS handshake performance using our
patched Go \texttt{crypto/tls} stack on an Intel i9-12900, built on
Go~1.26. The benchmark performs complete TLS~1.3 handshakes over
loopback for five scenarios: classical ECDSA~P-256, MTC standalone
(16~leaves, 2~cosignatures), and MTC landmark at three subtree sizes
(16, 1024, 4096~leaves). All scenarios use ECDSA~P-256 as the entity
key; post-quantum entity key overhead is projected analytically.

Table~\ref{tab:empirical} summarizes the results.

\begin{table*}[t]
\centering
\caption{Empirical TLS~1.3 Handshake Measurements (ECDSA P-256 entity key, loopback)}
\label{tab:empirical}
\small
\begin{tabular}{@{}p{4.2cm}rrrrr@{}}
\toprule
\textbf{Scenario} & \textbf{Cert (B)} & \textbf{Proof (B)}
    & \textbf{Verify} & \textbf{CertVerify}
    & \textbf{Handshake} \\
\midrule
\rowcolor{lightgrayrow}
Classical ECDSA P-256
    & 342 & - & 24.2~$\mu$s\textsuperscript{a}
    & 24.2~$\mu$s & 678~$\mu$s \\
MTC Standalone (16, 2 cosigs)
    & 590 & 128 & 1.07~$\mu$s
    & 24.2~$\mu$s & 612~$\mu$s \\
\rowcolor{lightgrayrow}
MTC Landmark (16 leaves)
    & 425 & 128 & 877~ns
    & 25.0~$\mu$s & 563~$\mu$s \\
MTC Landmark (1024 leaves)
    & 619 & 320 & 1.63~$\mu$s
    & 23.0~$\mu$s & 590~$\mu$s \\
\rowcolor{lightgrayrow}
MTC Landmark (4096 leaves)
    & 683 & 384 & 1.92~$\mu$s
    & 22.8~$\mu$s & 715~$\mu$s \\
\bottomrule
\end{tabular}

\vspace{0.4em}
\footnotesize
\textsuperscript{a}Classical verification is ECDSA signature
verification, not MTC proof evaluation. Averaged over 100--10,000
iterations. Handshake times over loopback, excluding network latency.
\end{table*}

MTC landmark verification completes in under 2~$\mu$s across all
subtree sizes -- 12--28$\times$ faster than ECDSA~P-256 signature
verification (24~$\mu$s). Entry hash computation is constant
($\sim$290~ns) regardless of subtree size; proof evaluation scales
logarithmically with the number of leaves, confirming the expected
$O(\log n)$ behaviour. Handshake times for MTC landmark
(563--715~$\mu$s) are comparable to classical ECDSA (678~$\mu$s),
confirming that MTC does not introduce measurable overhead.

\paragraph{Projected PQ comparison.}
With ML-DSA-65 as the entity key, certificate size increases by
approximately 1,887~bytes (the public key difference), and
\texttt{CertificateVerify} cost rises from $\sim$24~$\mu$s to
$\sim$120--200~$\mu$s~\cite{oqs-bench}. MTC proof verification is
unaffected---it operates on the entry hash, not the public key.
A PQ~X.509 chain verification (4~ML-DSA-65 signature verifications
at $\sim$150~$\mu$s each) would cost $\sim$600~$\mu$s, compared to
$\sim$1.9~$\mu$s for MTC landmark-approximately 300$\times$ faster.
Empirical validation with ML-DSA-65 entity keys is planned as future
work.

\subsection{Verification Cost}
\label{sec:verif-perf}

MTC landmark verification involves only SHA-256 hash computations.
For a subtree of size $n$, the cost is:
\begin{equation}
\text{Cost}_{\text{landmark}} = \lceil \log_2 n \rceil \times
    T_{\text{SHA-256}} + T_{\text{entry\_hash}}
\end{equation}

Our empirical measurements (Table~\ref{tab:empirical}) confirm this
scaling: verification grows from 877~ns at 16~leaves to 1.92~$\mu$s
at 4,096~leaves, consistent with logarithmic growth. Compared to
ECDSA~P-256 verification (24~$\mu$s measured), MTC landmark is
12--28$\times$ faster. Against a projected PQ~X.509 chain
(4~ML-DSA-65 verifications at $\sim$150~$\mu$s
each~\cite{oqs-bench}), landmark verification is approximately
300$\times$ faster.

\subsection{Relying Party State}

Table~\ref{tab:rp-state} shows the per-node state required to support landmark verification.

\begin{table*}[t]
\centering
\caption{Relying Party State Cost}
\label{tab:rp-state}
\small
\begin{tabular}{@{}lrr@{}}
\toprule
\textbf{Environment} & \textbf{Per CA} & \textbf{Total} \\
\midrule
\rowcolor{lightgrayrow}
K8s (1 CA, 1h landmarks) & 1,600~B & 1.6~KB \\
K8s (10 CAs) & 1,600~B & 16~KB \\
\rowcolor{lightgrayrow}
5G/6G (1 NRF, 10min landmarks) & 9,280~B & 9.3~KB \\
5G/6G (5 PLMNs roaming) & 9,280~B & 46.4~KB \\
\bottomrule
\end{tabular}
\end{table*}

Even in the most demanding roaming scenario with 5~partner PLMNs, the total landmark state is under 47~KB.

\section{Comparative Analysis}
\label{sec:comparison}

\subsection{MTC vs.\ Classical X.509 PKI}

Table~\ref{tab:comparison-x509} compares MTC with traditional X.509 across key dimensions.

\begin{table}[t]
\centering
\caption{MTC vs.\ Classical X.509 PKI}
\label{tab:comparison-x509}
\small
\begin{tabular}{@{}p{2.1cm}p{2.2cm}p{2.7cm}@{}}
\toprule
\textbf{Dimension} & \textbf{X.509} & \textbf{MTC} \\
\midrule
\rowcolor{lightgrayrow}
Trust anchor & Root CA cert & Subtree hash (32~B) \\
Verification & Sig.\ check per cert & Hash comp.\ (landmark) \\
\rowcolor{lightgrayrow}
Revocation & CRL/OCSP (network) & Index ranges (local) \\
Transparency & Optional CT & Built-in \\
\rowcolor{lightgrayrow}
PQ readiness & Requires PQ sigs & Landmark: sig-free \\
Infra.\ cost & Simple CA & CA + witnesses + mirrors \\
\bottomrule
\end{tabular}
\end{table}

MTC introduces additional infrastructure (witnesses, mirrors, landmark distribution) but provides transparency and PQ readiness that classical PKI lacks. In private deployments where the CA is already trusted, MTC's main advantage is bandwidth reduction and PQ readiness.

\subsection{MTC vs.\ Identity-Based PKI and TLS}

~\cite{rathi2025pqibtls} propose IBE-based PKI and TLS for 5G, eliminating certificates and signatures entirely. 
Table~\ref{tab:comparison-ibe} compares the approaches. The comparison highlights that IBE-TLS, does offer the least handshake overhead. It is worth noting, however, that IBE is more experimental and prone to single point failures, though ~\cite{rathi2025pqibtls} tries to mitigate that via use of threshold PKGs.

\begin{table*}[t]
\centering
\caption{Comparison of MTC-TLS and IBE-based TLS~\cite{rathi2025pqibtls}}
\label{tab:comparison-ibe}
\small
\begin{tabular}{@{}lll@{}}
\toprule
\textbf{Dimension} & \textbf{IBE-TLS} & \textbf{MTC-TLS} \\
\midrule

\rowcolor{lightgrayrow}
Trust anchor 
& Threshold PKG 
& MTCA with cosigners \\

Authentication mechanism 
& Identity-based (implicit authentication) 
& X.509-based MTC certificate (explicit authentication) \\

\rowcolor{lightgrayrow}
Authentication overhead 
& $\sim$5~KB 
& $\sim$6.2KB (landmark, ML-DSA public key, signature) \textsuperscript{$\ddagger$} \\

Transparency 
& Not supported 
& Built-in via Merkle tree log \\

\rowcolor{lightgrayrow}
Single point of failure 
& PKG compromise 
& CA compromise (mitigated by witnesses) \\

Revocation 
& Limited / underdeveloped 
& Efficient via index ranges \\

\rowcolor{lightgrayrow}
TLS compatibility 
& Requires modified handshake 
& Fully compatible with TLS~1.3 \\

Certificate chain 
& Not required 
& Depth 1 (leaf only) \\

\rowcolor{lightgrayrow}
Deployment complexity 
& PKG setup and key extraction 
& CA, witnesses, and landmark management \\

\bottomrule
\multicolumn{3}{@{}l}{\footnotesize\textsuperscript{$\dagger$}The MTC authentication overhead accounts for the landmark certificate,  signature public key and CertificateVerify message of TLS.}

\end{tabular}
\end{table*}
MTC's witness/mirror infrastructure provides cryptographic assurance that the CA has not mis-issued certificates, a property that IBE lacks.

Furthermore, MTC uses \emph{unmodified} X.509 certificate structures (the \texttt{signatureValue} field is reinterpreted), enabling backward compatibility with existing TLS implementations that can be upgraded incrementally.

\section{Challenges and Limitations}
\label{sec:challenges}

\subsubsection{Architectural Complexity vs.\ Marginal Gain in Private Deployments}

MTC introduces three new infrastructure components-issuance log, witness cosigners, and log mirror-on top of the CA that a traditional PKI already requires. In a private 5G Core where a single operator controls all NFs, the transparency guarantees that MTC provides (append-only logging, independent cosigner verification, monitor auditing) solve a problem that may not exist: the operator already trusts its own CA. Thus, it is upto the operator whether certificate logging is required or not. A mis-issuance in a private deployment is an internal misconfiguration, not an adversarial attack, and simpler mechanisms-short-lived self-signed certificates with SPIFFE IDs, or a lightweight internal CA like \texttt{step-ca} with automatic rotation-achieve comparable security with far less infrastructure. The bandwidth reduction from landmark certificates is real, but only matters in post-quantum settings; with classical ECDSA, MTC landmark certificates (754\,B) are actually \emph{larger} than a traditional Ed25519 certificate chain ($\sim$288\,B). Operators must weigh whether the PQ-readiness justifies the operational overhead before MTC's PQ threat model becomes relevant to their deployment.

\paragraph{\textbf{Component management}}
Additionally, an MTC cloud-native deployment requires several components: an MTCA pod, at least 2 cosigner pods, tree server/mirror pods, services for landmark distribution (optional), and certificate renewal jobs. This is an addition of several distinct workload types, each with its own lifecycle and management criteria. By contrast, a traditional PKI deployment (e.g., cert-manager with a self-signed Issuer) is a single controller pod. The operational cost is not in compute but in the cognitive load of understanding, debugging, and maintaining the interactions between components. For example, a log mirror desyncing with the CA logs could fail to catch CA misbehaviour, or provide ambiguity to the relying parties inspecting it.

\subsubsection{Cross-PLMN Trust and Cosigner Scaling}

5G/6G's roaming scenarios are enabled by the establishment of cross-PLMN trust. In an MTC-based deployment, each PLMN would operate their issuance log, and set of cosigners. For a relying party in PLMN~A to accept a certificate from PLMN~B, it must trust PLMN~B's cosigners-or a subset of them that satisfies its acceptance policy . With $n$ roaming partners, the relying party must track $n$ separate cosigner sets, $n$ landmark sequences, and $n$ revocation lists. The SEPP can mediate this exchange during N32-c peering setup, but the protocol for bootstrapping cross-PLMN cosigner trust is undefined. Defining how cosigner public keys, acceptance policies, and landmark hashes are negotiated between PLMNs is an \textit{open problem}.

\subsubsection{Standardization Gap}

Telecom networks are usually always defined through standards, which allow for cross-operator interoperability, network operational and security guarantees, etc. Conforming to standards establishes trust, and boosts an organization's credibility. Telecom standards are typically written by 3GPP, and the O-RAN alliance, which carry out dedicated discussions, feasibility studies, and follow a rigorous methodology before adopting a new standard. Currently, however, Merkle Tree Certificates is not referenced by any 3GPP study as per our knowledge. The IETF draft targets WebPKI deployments, and not inherently telecom infrastructure. Thus, a wide adoption of MTC would require (not an exhaustive list):
\begin{itemize}[nosep]
    \item A 3GPP SA3 study item evaluating MTC for SBI authentication,
    \item Extensions to TLS 1.3 used in SBA,
    \item Upgrades to zero-trust requirements,
    \item Agreement on cosigner governance across operators.
\end{itemize}
Each of these is a multi-year effort. Until then, MTC can only be deployed as a vendor extension, which limits interoperability with NFs from other vendors or 3GPP-conformant implementations that do not recognize \texttt{id-alg-mtcProof}.

\subsubsection{Specification Maturity}

The IETF draft is at version~02 and still evolving. Between versions~01 and~02, the cosigner signature format changed, the landmark allocation procedure was revised, and the certificate format gained new fields. Building production infrastructure on a draft that may change materially before reaching RFC status carries risk: our implementation may require rework if the wire format changes, and certificates issued under the current format may become unparseable by future implementations. The PLANTS working group has not yet reached consensus on several open issues, including whether the log entry format should include additional metadata.

\subsubsection{Landmark Staleness and Verification Fallback}

Landmark verification only works when the relying party holds the subtree hash that covers the certificate's log index. If the relying party's landmark set is stale-because the DaemonSet pod on its node crashed, or the tree server is lagging, or the NF pod was recently scheduled on a fresh node-it cannot verify landmark certificates from peers whose entries fall in newer subtrees. The fallback is standalone verification, which requires evaluating cosignatures. But if the relying party also lacks the cosigner public keys (same staleness problem), verification fails entirely. The system degrades gracefully only if \emph{at least one} trust path (landmark or cosigner) is available. Coordinating the freshness of landmarks, cosigner keys, and revocation lists across hundreds of NF pods on dozens of nodes is an operational challenge that grows with cluster size.

\subsubsection{Log Pruning and Entry Availability}

The IETF draft permits CAs to prune expired entries from the log (Section~5.6.1). Once pruned, an entry's certificate can no longer be verified against the log-the leaf is gone, and inclusion proofs that reference it become invalid. The draft recommends preemptive revocation of pruned indices (Section~7.5), but the coordination between pruning and revocation distribution is left to the deployment. If the CA prunes an entry before all relying parties have revoked that index, a relying party could receive a certificate for a pruned-but-not-yet-revoked index and fail to verify it-not because the certificate is bad, but because the proof infrastructure no longer supports it. The failure mode is a false reject, not a false accept, so it is safe but disruptive.


\section{Future Work}

\subsection{Decentralized PKI with MTC for Cloud-Native}
The distributed trust model introduced by MTC cosigners and mirrors
could be extended toward a fully decentralized PKI, where multiple
independent operators co-manage a shared issuance log without a single root of trust.

\subsection{MTC integration with CMPv2}
CMPv2~\cite{cmp-rfc9480} is the latest revision of the Certificate
Management Protocol (originally RFC~4210), widely used in telecom for
automated certificate lifecycle management across Radio Access Networks
and core infrastructure. CMPv2 defines interactions between End Entities,
Registration Authorities (RAs), and Certificate Authorities for
certificate enrolment, renewal, revocation, and key update, and is
referenced by 3GPP for certificate provisioning in 5G networks
(TS~33.310).

MTC could integrate with CMPv2 at two levels. First, the MTCA could
serve as a backend CA behind a CMPv2 RA: the RA accepts standard CMPv2
initialization and certification requests (\texttt{ir}/\texttt{cr})
from network elements, forwards the public key and identity to the
MTCA's issuance API, and returns the MTC certificate in the CMPv2
response (\texttt{ip}/\texttt{cp}). This preserves the existing CMPv2
enrolment workflow that operators have deployed across their RAN and
transport networks. Second, CMPv2's revocation request (\texttt{rr})
could be mapped to MTC's revocation-by-index mechanism, where the RA
translates a certificate identifier to a log index and calls the MTCA's
\texttt{/revoke} endpoint. The transparency and append-only logging
properties of MTC would add auditability to the CMPv2 lifecycle that
the protocol currently lacks. Detailed protocol mapping between CMPv2
message types and MTCA APIs is left as future work.

\subsection{MTC-TLS Refinements: Eliminating CertificateVerify}
\label{sec:drop-certverify}

In the current MTC-TLS design, the server still sends a \texttt{CertificateVerify} message containing a digital signature over the handshake transcript. For landmark certificates, this is the \emph{only} signature in the entire authentication path. If the entity key is ML-DSA-65, this single signature is 3,309~bytes-larger than the entire MTC certificate.

A natural refinement (but rather less-sound) is to \textbf{eliminate the \texttt{CertificateVerify} message entirely} for MTC landmark certificates. The argument is as follows:

\begin{enumerate}[nosep]
    \item The MTC inclusion proof binds the public key to the Merkle tree via \texttt{SHA-256(SPKI)}.
    \item The landmark hash is pre-distributed to the relying party over a trusted channel.
    \item The certificate's short lifetime (e.g., 24~hours) limits the window of compromise.
    \item TLS~1.3's key schedule already incorporates the server's certificate into the handshake transcript hash, providing implicit authentication.
\end{enumerate}

However, dropping \texttt{CertificateVerify} removes the proof of private key possession. An adversary who obtains a valid MTC certificate (e.g., by compromising the log) could impersonate the server without knowing the private key. This trade-off is acceptable only in environments where:
\begin{itemize}[nosep]
    \item The issuance log is well-monitored (unauthorized entries are detected quickly).
    \item Certificate lifetimes are short (limiting impersonation windows).
    \item The deployment is private (reducing the attack surface).
\end{itemize}

This refinement would reduce the handshake by one full PQ signature (3,309~bytes for ML-DSA-65), making MTC-TLS authentication \emph{entirely hash-based} on the wire. 

\subsection{Satellite PKI and Non-Terrestrial Networks}
\label{sec:satellite}

Non-Terrestrial Networks (NTN), which was standardized in 3GPP Release 17 \cite{3gpp-38300, 3gpp-ntn-overview}, extends 5G connectivity beyond ground-based infrastructure (terrestrial) by integrating satellites, high altitude platforms (HAPS) and even drones into the network. This provides satellite access to smartphones and devices in areas lacking adequate terrestrial coverage, such as rural regions, oceans, forests, etc. It also solidifies 5G's goal of Massive-IoT, which is accomplished through LPWAN technologies like NB-IoT. Release 17 provides a link between NB-IoT and NTN, allowing for IoT literally everywhere. It introduces several modifications to counter large propagation delays, and Doppler shifts (which occur with LEO and MEO satellites). NTN can be categorized into 2 major designs: (i) transparent payload (where the satellite acts a relay/repeater). (ii) Regenerative payload, which performs base station functionalities, and links to the core network.
However, the NTN security must be guaranteed by the use of PKI, which becomes harder when long range networks are involved. 

\textbf{High latency:} Geostationary (GEO) satellites have round-trip
times of $\sim$600~ms. Low Earth Orbit (LEO) constellations have RTTs
of 30--80~ms depending on orbit altitude and routing
topology~\cite{3gpp-38300, 3gpp-ntn-overview}. Every additional
round-trip for OCSP or CRL retrieval compounds this latency. Several other modes could also thus rely on PSK-based authentication. Note: However, this effect is also amplified by the use of KEMs, which are interactive in nature, a \textit{Non Interactive Post-quantum key exchange} could alleviate major round trip times. This would make frequent rekeying much less expensive.

\textbf{Constrained bandwidth:} Feeder and service links have limited
capacity, especially for large PQ certificate chains that may exceed 17-20~KB per handshake. 

\textbf{Intermittent connectivity:} Satellites may lose ground station
contact during orbit, making real-time revocation checks impossible.

MTC addresses these challenges:
\begin{itemize}[nosep]
    \item \textbf{No revocation round-trips:} Revocation-by-index is
    checked locally against a pre-distributed list. Completely offline.
    \item \textbf{Compact certificates:} MTC landmark certificates
    carry approximately 2.9~KB on the wire (with ML-DSA-65 verification key),
    compared to $\sim$19.5~KB for a PQ X.509 chain-an 85\% reduction.
    \item \textbf{Offline verification:} Landmark hashes can be
    pre-loaded during ground station contacts and remain valid for the
    certificate lifetime.
    \item \textbf{Batch landmark updates:} Landmark hashes are 32~bytes
    each and can be bundled into periodic telemetry uplinks.
\end{itemize}

With 1-hour landmarks and 24-hour certificate lifetimes, each satellite
maintains roughly 25 active landmark hashes. For a 1,000-satellite LEO
constellation under a single CA, this amounts to
$1{,}000 \times 25 \times 32 = 800$~KB of total state-small enough
to distribute over a few telemetry frames.

\section{Conclusion}
\label{sec:conclusion}

In this paper, we have presented the design and analysis of Merkle Tree
Certificate-based PKI for Kubernetes control planes and cloud-native 5G
Core networks, building on the ongoing MTC specification at IETF. Our
key findings are:

\begin{enumerate}[nosep]
    \item \textbf{MTC integrates naturally with Kubernetes} via custom
    controllers that handle Merkle Tree certificate requests, trust
    bootstrapping from the existing X.509 infrastructure, and
    DaemonSet-based landmark distribution.
    \item \textbf{MTC can be introduced to cloud-native 5G Core} in a
    modular, decoupled fashion that minimally modifies existing Network
    Functions, while providing certificate transparency guarantees and
    reducing post-quantum mTLS bandwidth and authentication overhead by
    approximately 85\% compared to PQ X.509 chains.
    \item \textbf{Relying party state is trivially small}-under
    16~KB for 10~CAs in Kubernetes, under 47~KB for 5-PLMN roaming
    in 5G.
    \item \textbf{MTC landmark verification is hash-only}, requiring
    no public-key operations for certificate authentication, 12--28$\times$ faster than classical ECDSA verification (measured), and projected 200--400$\times$ faster than ML-DSA-65 chain verification.

\end{enumerate}
Looking forward, we aim to enhance the current architecture with
several optimizations-including the (possible) elimination of the
\texttt{CertificateVerify} message for landmark certificates, which
would make MTC-TLS authentication entirely hash-based on the wire.
Our proposed 6G-native architecture, where the NRF serves as the MTCA
and the SCP as witness cosigner, removes the need for a separate PKI
subsystem entirely; we aim to refine this design and work towards
informing 3GPP standardization efforts for native MTC support in
future core network specifications. MTC's compact, offline-verifiable
certificates also make it a strong candidate for satellite PKI and
Non-Terrestrial Networks (NTN), where bandwidth constraints and
intermittent connectivity make traditional PQ certificate chains and
online revocation checks impractical. We believe this work provides a
practical foundation for post-quantum PKI in private infrastructure,
complementing the WebPKI focus of the IETF PLANTS working group
specification.


\bibliographystyle{IEEEtran}
\bibliography{references}

\appendix

\section{\\Durable Logging, Revocation and Misbehaving CA}
The IETF draft highlights a subtlety regarding durable logging (Section~7.5): a misbehaving CA might construct a globally consistent log but refuse to make some entries available to monitors. In this scenario, consistency proofs between checkpoints pass, but the hidden entries cannot be audited. Relying parties whose cosigner policies do not require mirror cosignatures are vulnerable to this attack, because a consistency-checking witness only verifies the tree structure, not the availability of individual entries.

To mitigate this, the draft recommends two complementary mechanisms:
\begin{itemize}[nosep]
    \item \textbf{Mirror cosignatures}: Requiring at least one mirroring cosigner in the acceptance policy ensures that all entries have been verified as available before the cosignature is issued. In our deployment, the log mirror fulfills this role.
    \item \textbf{Preemptive index revocation}: When a relying party is first configured to trust a CA, it should revoke all indices from zero up to the first available unexpired entry. This revocation is periodically updated as log pruning advances the minimum index. Even if a hidden entry has not yet expired, the relying party will not accept it.
\end{itemize}

\begin{figure}[h]
    \centering
    \fbox{\includegraphics[width=1.0\columnwidth]{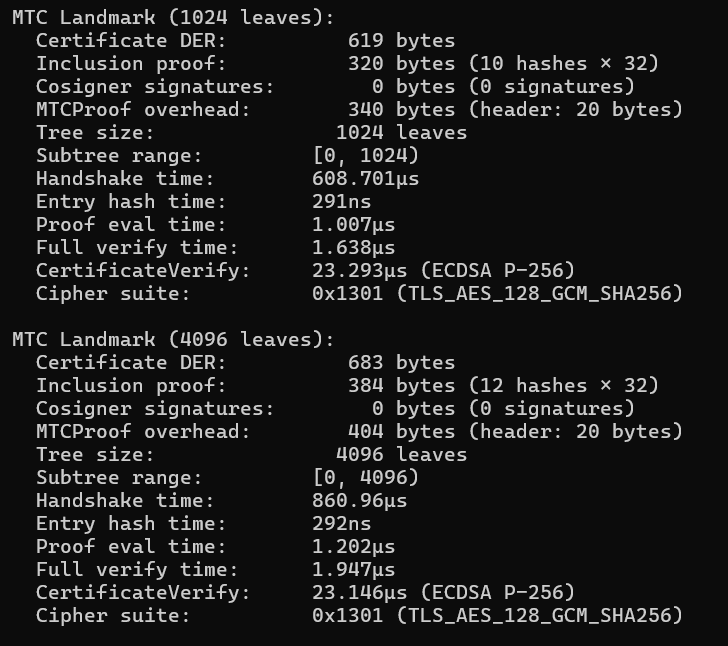}}
    \caption{Landmark certificates with varying number of leaves}
    \label{fig:landmark}
\end{figure}

\newpage
\section{MTC-PKI evaluation logs}
\begin{figure*}
    \centering

    \begin{subfigure}{\textwidth}
        \centering
\begin{tcolorbox}[boxrule=0.4pt, colback=white, colframe=black!40,
                  sharp corners, boxsep=0pt, left=2pt, right=2pt,
                  top=2pt, bottom=2pt]
    \includegraphics[width=\linewidth]{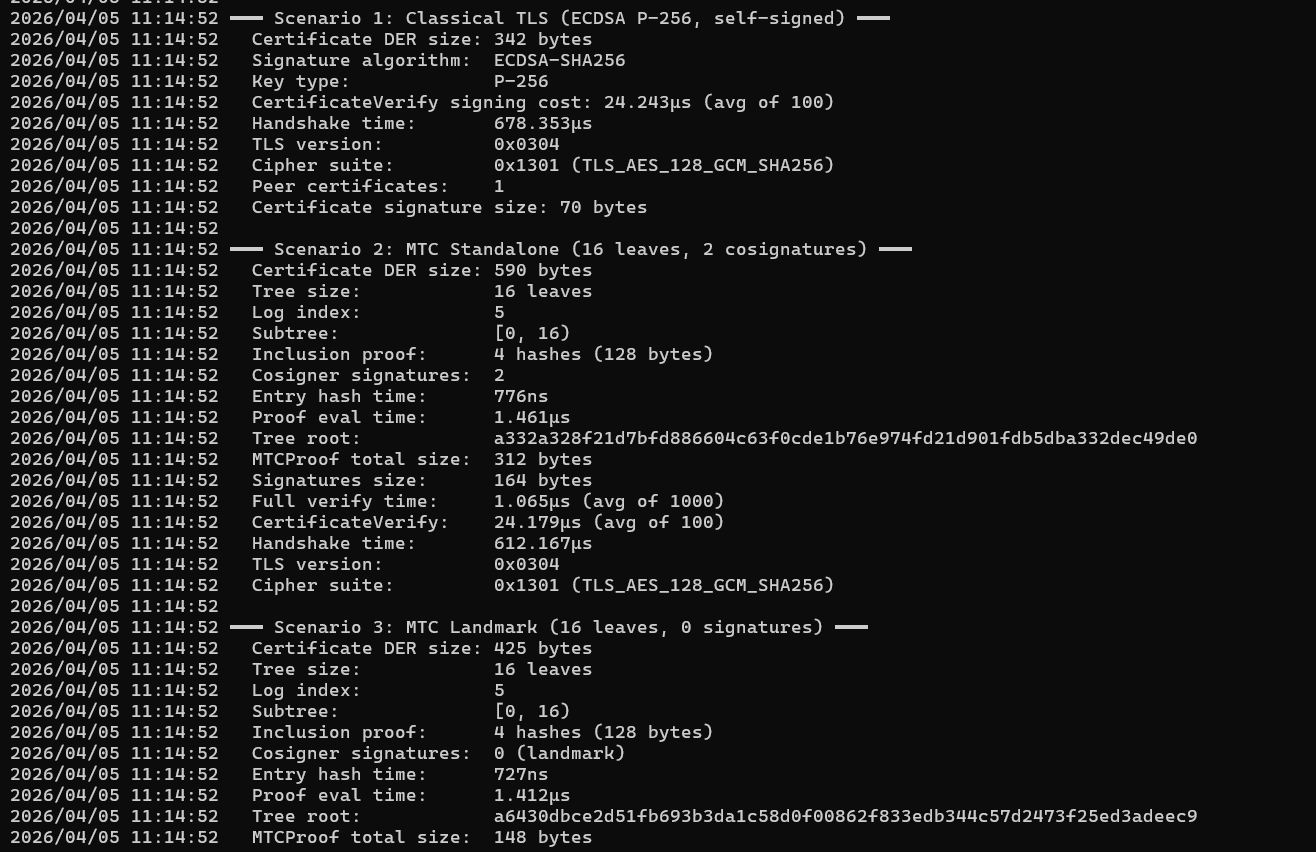}
\end{tcolorbox}

        \caption{Results comparing ECDSA P-256 with MTC (Standalone and Landmark modes)}
        \label{fig:first_comparison}
    \end{subfigure}

    \vspace{0.5cm}

    \begin{subfigure}{0.8\textwidth}
        \centering
        \begin{tcolorbox}[boxrule=0.4pt, colback=white, colframe=black!40,
                          sharp corners, boxsep=0pt, left=2pt, right=2pt,
                          top=2pt, bottom=2pt]
            \includegraphics[width=\linewidth]{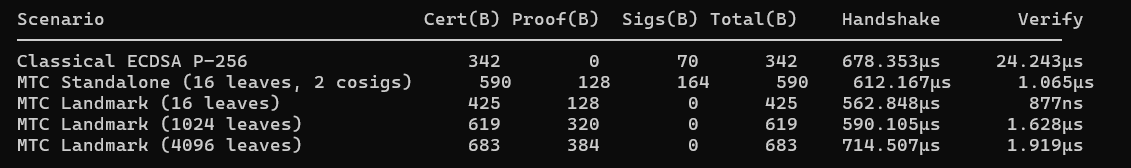}
        \end{tcolorbox}
        \caption{Tabulated results}
        \label{fig:res_table}
    \end{subfigure}

    \vspace{0.3cm}

    \begin{subfigure}{0.8\textwidth}
        \centering
        \begin{tcolorbox}[boxrule=0.4pt, colback=white, colframe=black!40,
                          sharp corners, boxsep=0pt, left=2pt, right=2pt,
                          top=2pt, bottom=2pt]
            \includegraphics[width=\linewidth]{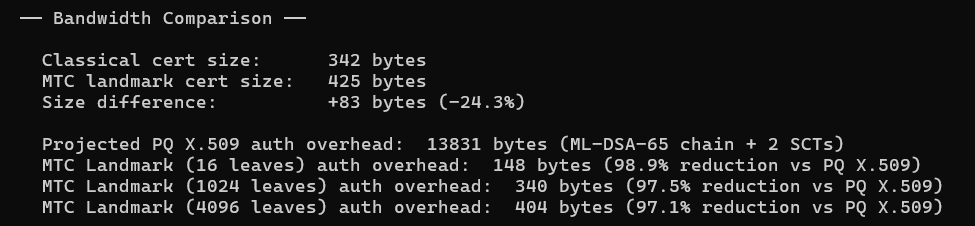}
        \end{tcolorbox}
        \caption{Expected reductions with MTC}
        \label{fig:reductions}
    \end{subfigure}

    \caption{MTC-PKI evaluation logs}
\end{figure*}

\balance

\end{document}